\pdfoutput=1

\newcommand{\akari}{\textit{AKARI}}
\newcommand{\wise}{\textit{WISE}}
\newcommand{\iras}{\textit{IRAS}}

\newcommand{\oiilong}{\hbox{[O\sc ii] 3726\AA\ 3729\AA}}
\newcommand{\oii}{\hbox{[O\sc ii]}}
\newcommand{\neiii}{\hbox{[Ne\sc iii]}}
\newcommand{\neiiilong}{\hbox{[Ne\sc iii] 3869\AA}}

\newcommand{\oiiiblong}{\hbox{[O\sc iii] 5007\AA}}
\newcommand{\oiiilong}{\hbox{[O\sc iii] 4959\AA\ 5007\AA}}
\newcommand{\oiii}{\hbox{[O\sc iii]}}
\newcommand{\nii}{\hbox{[N\sc ii]}}
\newcommand{\niilong}{\hbox{[N\sc ii] 6548\AA\ 6583\AA}}

\newcommand{\niiblong}{\hbox{[N\sc ii] 6583\AA}}

\newcommand{\sii}{\hbox{[S\sc ii]}}
\newcommand{\siilong}{\hbox{[S\sc ii] 6716\AA\ 6731\AA}}

\newcommand{\ha}{H$\alpha$}
\newcommand{\hb}{H$\beta$}
\newcommand{\hii}{\hbox{H\sc ii}}

\newcommand{\ms}{M$_{\odot}$}

\newcommand{\ls}{L$_{\odot}$}
\newcommand{\kms}{km~s$^{-1}$}
\newcommand{\fluxcgs}{erg~s$^{-1}$~cm$^{-2}$}
\newcommand{\lumcgs}{erg~s$^{-1}$}
\newcommand{\lumir}{$L_{\rm IR}$}

\newcommand{\sfrunit}{M$_{\odot}$ yr$^{-1}$}

\RequirePackage{fix-cm} 

\documentclass[otwocolumn]{pasj01}
\Received{$\langle$reception date$\rangle$}
\Accepted{$\langle$acception date$\rangle$}
\Published{$\langle$publication date$\rangle$}

\usepackage{graphicx}
\usepackage{ulem}

\begin{document}

\title{Discovery of a strong ionized-gas outflow in an \textit{\textbf{AKARI}}-selected Ultra-luminous Infrared Galaxy at \textit{\textbf{z}} = 0.5}
\author{
Xiaoyang \textsc{Chen}\altaffilmark{1}, 
Masayuki \textsc{Akiyama}\altaffilmark{1}, 
Hirofumi \textsc{Noda}\altaffilmark{2}, 
\textsc{Abdurro'uf}\altaffilmark{1,3}, 
Yoshiki \textsc{Toba}\altaffilmark{4,5}, 
Issei \textsc{Yamamura}\altaffilmark{6,7}, 
Toshihiro \textsc{Kawaguchi}\altaffilmark{8}, 
Mitsuru \textsc{Kokubo}\altaffilmark{1}, 
Kohei \textsc{Ichikawa}\altaffilmark{1,9} }

\altaffiltext{1}{Astronomical Institute, Tohoku University, 6-3 Aramaki, Aoba-ku, Sendai, Miyagi 980-8578, Japan}
\altaffiltext{2}{Department of Earth and Space Science, Graduate School of Science, Osaka University, 1-1 Machikaneyama-cho, Toyonaka-shi, Osaka 560-0043, Japan} 
\altaffiltext{3}{Academia Sinica Institute of Astronomy \& Astrophysics, 11F of Astronomy-Mathematics Building, AS\,/\,NTU, No.1, Section 4, Roosevelt Road, Taipei 10617, Taiwan, R.O.C.}
\altaffiltext{4}{Department of Astronomy, Kyoto University, Kitashirakawa-Oiwake-cho, Sakyo-ku, Kyoto 606-8502, Japan}
\altaffiltext{5}{Research Center for Space and Cosmic Evolution, Ehime University, 2-5 Bunkyo-cho, Matsuyama, Ehime 790-8577, Japan}
\altaffiltext{6}{Institute of Space and Astronautical Science, JAXA, 3-1-1 Yoshinodai, Chuo-ku, Sagamihara, Kanagawa 252-5210, Japan}
\altaffiltext{7}{Department of Space and Astronautical Science, SOKENDAI, 3-1-1 Yoshinodai, Chuo-ku, Sagamihara, Kanagawa 252-5210, Japan}
\altaffiltext{8}{Department of Economics, Management and Information Science, Onomichi City University, Hisayamada 1600-2, Onomichi, Hiroshima 722-8506, Japan}
\altaffiltext{9}{Frontier Research Institute for Interdisciplinary Sciences, Tohoku University, 6-3 Aramaki, Aoba-ku, Sendai, Miyagi 980-8578, Japan}

\email{xy.chen@astr.tohoku.ac.jp}


\maketitle

\begin{abstract}
In order to construct a sample of ultra-luminous infrared galaxies (ULIRGs, with infrared luminosity, $L_{\rm IR} > 10^{12}$ L$_{\odot}$) at $0.5<z<1$, we are conducting an optical follow-up program for bright 90-$\mu$m FIR sources 
with a faint optical ($i<20$ mag) counterpart 
selected in the \akari\ Far-Infrared Surveyor (FIS) Bright Source catalog (Ver.2). 
AKARI-FIS-V2 J0916248+073034, identified as a ULIRG at $z=0.49$ in the spectroscopic follow-up observation, indicates signatures of an extremely strong outflow in its emission line profiles. 
Its \hbox{[O\sc iii] 5007\AA} emission line shows FWHM of $1830$ \kms\ and velocity shift of $-770$ \kms\ in relative to the stellar absorption lines. 
Furthermore, low-ionization \hbox{[O\sc ii] 3726\AA\ 3729\AA} doublet also shows large FWHM of $910$ \kms\ and velocity shift of $-380$ \kms.
After the removal of an unresolved nuclear component, the long-slit spectroscopy 2D image possibly shows that the outflow extends to radius of 4 kpc. 
The mass outflow and energy ejection rates are estimated to be $500$ \sfrunit\ and $4\times10^{44}$ \lumcgs, respectively, which imply that the outflow is among the most powerful ones observed in ULIRGs and QSOs at $0.3<z<1.6$. 
The co-existence of the strong outflow and intense star formation (star formation rate of 990 \sfrunit) indicates that the feedback of the strong outflow has not severely affect the star-forming region of the galaxy. 
\end{abstract}

\section{Introduction} 
\label{sec:introduction}
Ultra-luminous infrared galaxies (ULIRGs, with infrared luminosity, $L_{\rm IR} > 10^{12}$ L$_{\odot}$) are a population of 
the most intensely star-forming galaxies in the local universe, with star formation rates (SFR) of $10^2$--$10^3$ \sfrunit \citep{Sanders1996, Robinson2000}. 
They are thought to represent rapidly growing phase of massive galaxies before quenching of their star formation. 
In a widely accepted scenario of massive galaxy formation, a ULIRG is thought to be formed after merging of two gas-rich disk galaxies. 
An intense star formation in the nucleus is triggered by a rapid feeding of gas into the central region. This star formation results in a formation of dust particles, which, in turn, absorb rest-frame optical and ultraviolet (UV) emission from young blue stars, and re-radiate at far-infrared (FIR)\,/\,sub-millimeter wavelengths. 
Following the feeding of gas into the nuclear region, an active galactic nucleus (AGN) emerges \citep{Sanders1988, Hopkins2008}. 
As AGN heats the dusty materials around the supermassive blackhole (SMBH), 
the spectral energy distribution (SED) of the ULIRG shows a clear mid-infrared (MIR) bump since the temperature of dust heated by AGN is higher than that heated by the star formation \citep{Imanishi2007, Alonso2012, RodriguezL2013, Ichikawa2014}. 
The vigorous starburst and\,/\,or AGN in a ULIRG can induce a strong outflowing wind, which would blow out the gas and dust and terminate the activity of the galaxy. 
It is commonly accepted that a powerful AGN is required to drive outflows to high velocity, e.g., $v_{\rm max} \ge 500$ \kms\ ($v_{\rm max}=| v_{\rm shift} | + v_{\rm dispersion}$, \cite{Rupke2005, Westmoquette2012, Arribas2014, Harrison2014}). 

Recently, signatures of such outflowing gas in multi-phase have been found in various ULIRGs 
(neutral, \cite{Rupke2011}, \cite{Perna2015}; 
ionized, \cite{Soto2012}, \cite{RodriguezZ2013}; 
and molecular, \cite{Veilleux2013}, \cite{Saito2017}). 
Those observations support the idea that outflow plays an important role in the transition of a ULIRG from an extreme starburst to a quiescent galaxy. 
One widely employed diagnostic to identify the ionized outflowing gas is the broad \oiiiblong\ emission line profiles \citep{Christopoulou1997, Tadhunter2001, Zamanov2002}. 
As a forbidden (collisionally excited) transition, the \oiii\ line is a good tracer of the kinematics of 
gas in \hii\ regions formed by hot, young stars in a massive starburst, 
or gas in the narrow line region (NLR) formed by an AGN \citep{Osterbrock2006}, which is found to be extended over parsecs to several kiloparsecs by the spatially resolved long-slit and integral-field spectroscopy observations (e.g., \cite{Westmoquette2012, Liu2013, Harrison2012}). 
Compared to the \oiii\ line with relatively high ionization potential (IP, 35.12 eV), the \oiilong\ doublet with lower IP (13.62 eV) can trace ionized gas with lower ionization degree, which shows more extended structure than gas traced by \oiii\ line \citep{Collins2009}. 
Several works reported that \oii\ emission line also shows outflow signatures, although it usually does not display the extremely broad feature seen in \oiii\ emission line \citep{Zakamska2014, Perna2015, Toba2017}. 
Broad \oii\ line profiles possibly imply that the ionized outflow could 
couple to the interstellar medium (ISM) at larger radius, i.e., over galaxy-wide scales ($\sim10$ kpc). 

In this paper we report a discovery of one ULIRG selected from \akari\ 90-$\mu$m FIR survey, AKARI-FIS-V2 J0916248+073034 (hereafter J0916a), which indicates signatures of an extremely strong outflow in both \oiii\ and \oii\ emission line profiles but possibly with low AGN contribution to its bolometric luminosity.  The paper is organized as follows. 
We present the photometric and spectroscopic observations and data reduction in Section \ref{sec:data_reduction}. 
The methods and results of optical spectral analyses and multi-band SED fitting are shown in Section \ref{sec:spectral_analyses} and Section \ref{sec:sed_analyses}, respectively. 
We discuss the properties of the extreme outflow in Section \ref{sec:discuss}. 
Section \ref{sec:conclusion} summarizes the conclusion. 
Throughout the paper, the rest frame wavelengths are given in the air. We adopt the cosmological parameters $H_0=$ 70 \kms\,Mpc$^{-1}$, $\Omega_m=0.3$ and $\Omega_{\Lambda}=0.7$.


\section{Observation and Data Reduction} 
\label{sec:data_reduction}

In order to construct a sample of ULIRGs at intermediate redshifts ($0.5<z<1$), we are conducting an optical follow-up program for 90-$\mu$m FIR sources in the \akari\ Far-Infrared Surveyor (FIS) Bright Source catalog (Ver.2, hereafter FISBSCv2, Yamamura et al., in prep.) by utilizing the Sloan Digital Sky Survey (SDSS) optical imaging data. 
We focus on optically-faint sources, which are expected to be at relatively high redshifts ($z\sim1$) and luminous in the IR wavelength range. 
The detection limit of FISBSCv2 catalog reaches 0.44 Jy at 90 $\mu$m, which is deeper than the previous FIR all-sky survey by the Infrared Astronomical Satellite (\iras). 
Furthermore, thanks to the smaller PSF of \akari\ (spatial resolution of FWHM $=1$--$1.5'$, \cite{Doi2015}) than \iras, 
we can achieve reliable identifications even with faint ($i\sim20$) optical counterparts. 
Additionally, recently released Wide-field Infrared Survey Explorer (\wise) all-sky survey catalog \citep{Wright2010} in the MIR wavelength range (3.4 (W1), 4.6 (W2), 12 (W3), and 22 (W4) $\mu$m) with a spatial resolution of FWHM $=6.1$--$12.0''$, is useful to narrow down the positional uncertainty of the \akari\ FIR sources. 
The detection limit of the \wise\ survey is significantly deeper (1 mJy at 12 $\mu$m) than the \akari\ FIR catalog for sources with typical IR SED (e.g., \cite{Polletta2007}).
Therefore, the \akari\ FIR survey with the \wise\ pinpointing and SDSS optical photometric data is unique in identifying rare luminous FIR galaxies with faint optical counterpart.
The cross-matched sample covers $\sim7000$ deg$^2$, which is 10 times wider than the \textit{Herschel} Astrophysical Terahertz Large Area Survey (H-ATLAS) FIR survey at the similar depth \citep{Smith2017}.

We started spectroscopic follow-up of the optically-faint FIR sources using FOCAS on the Subaru telescope. Eight objects are observed in a service program in S17A (S17A0216S, PI:Masayuki Akiyama).  
For each object, the long slit spectra were obtained from two exposures, with integration time of 420 seconds per exposure. 
The 300B grism with SY47 filter were used. The configuration provides spectral resolution of $\rm R=500$ at 5500 \AA\ with slit width of 0.5$''$ and covers 4600--9200 \AA\ in the first-order of SY47. 
The data is reduced through bias-subtraction, flat-fielding, cosmic rays removal and wavelength calibration, before extraction of an individual spectrum using the IRAF package \texttt{focasred}. 
The wavelength calibration accuracy and spectral resolution measured using the positions and widths of the night-sky emission lines are 1 \AA\ and 10.6 \AA\ (FWHM), respectively, in the entire spectrum. 
The flux calibration is done with the spectra of standard stars, Feige34 and GD153, taken after the observations of the main targets. 
The atmospheric reddening was corrected using the Maunakea mean atmospheric extinction curve from \citet{Buton2013}. 
We also corrected for the atmospheric oxygen and water vapor absorptions by comparing the observed spectra of standard stars with those archived in the \texttt{CALSPEC} database\footnote{http://www.stsci.edu/hst/observatory/crds/calspec.html}. 
The slit-loss effect was estimated to be approximately 50\,\% by comparing the integrated flux to the SDSS photometry in the \textit{g}- and \textit{i}-bands, and corrected in the reduced spectrum. 
The seeing during the observation was 0.5$''$. 

\begin{figure}[!ht]
    \begin{center}
    \includegraphics[width=0.24\textwidth]{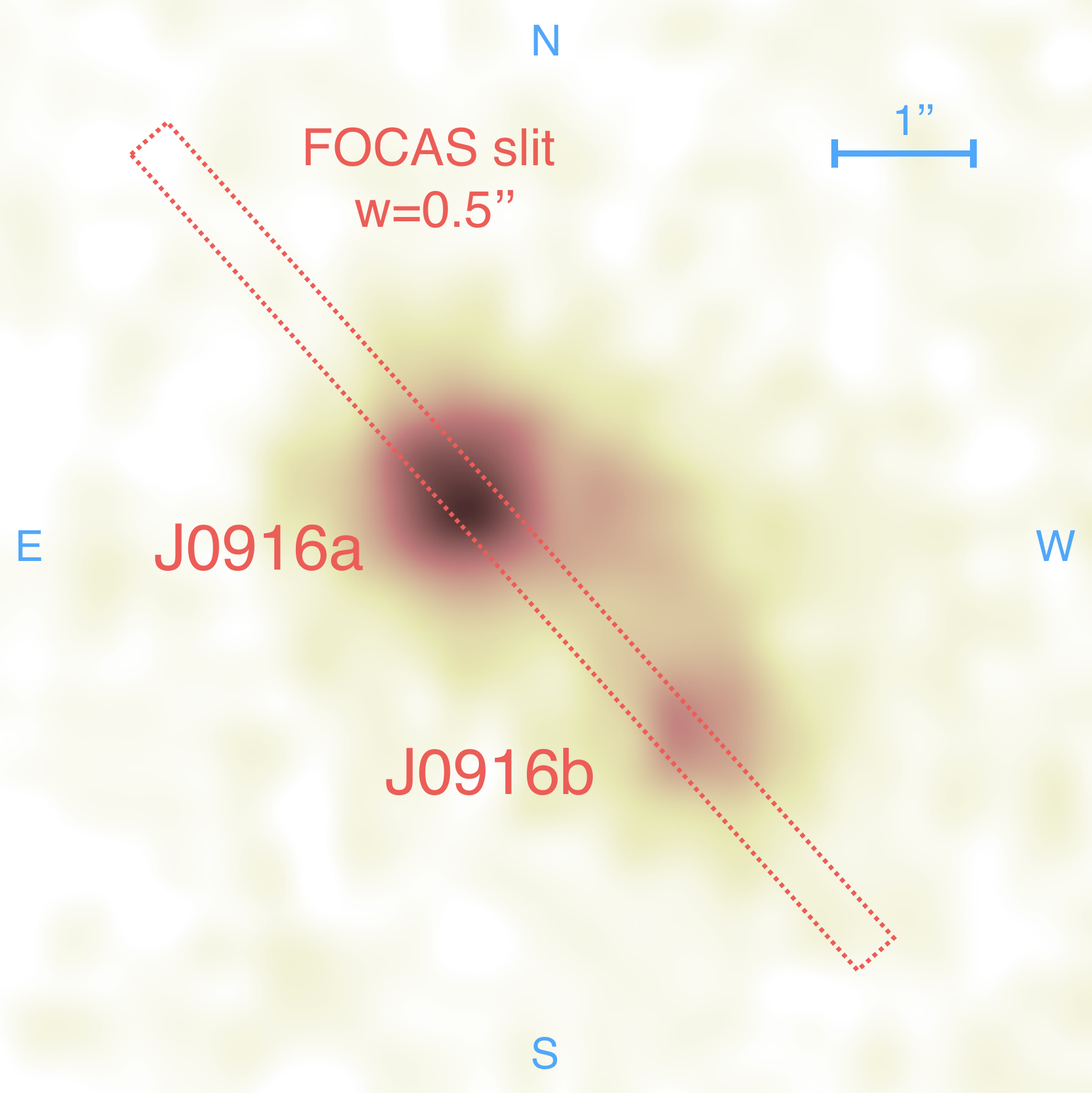}
    \end{center}
    \caption{Subaru\,/\,FOCAS image of J0916a and J0916b. 
    The image was taken without filter during acquisition. The wavelength range should be 4500\AA--9500\AA. 
    The location of the Subaru\,/\,FOCAS long-slit is shown with a red rectangle. }
    \label{fig:image_focas}
\end{figure}

All of the eight objects are identified as emission line galaxies at $0.3<z<0.6$. 
One of the objects, J0916a with \textit{i}-band magnitude of 20.29 mag shows emission lines at $z=0.49$ with features of a strong outflow.
Since the \ha-\nii\ complex of J0916a lies around 9800 \AA\ in the observed frame, we also examine the observed spectrum in the 9200--10000 \AA\ range after correcting for the contamination of the second-order spectrum, using the 
throughput of the first- and second-order of the 300B+SY47 setting\footnote{https://www.subarutelescope.org/Observing/Instruments/FOCAS/spec/\\efficiency.html} and the observed first-order spectrum below 5000 \AA.
The optical image of the object is shown in Figure \ref{fig:image_focas}. The object has a neighboring galaxy J0916b with \textit{i}-band magnitude of 21.43 mag. During the observation of this object the slit was aligned to the direction along the two galaxies. 

\begin{figure}[!ht]
    \begin{center}
    \includegraphics[width=0.48\textwidth]{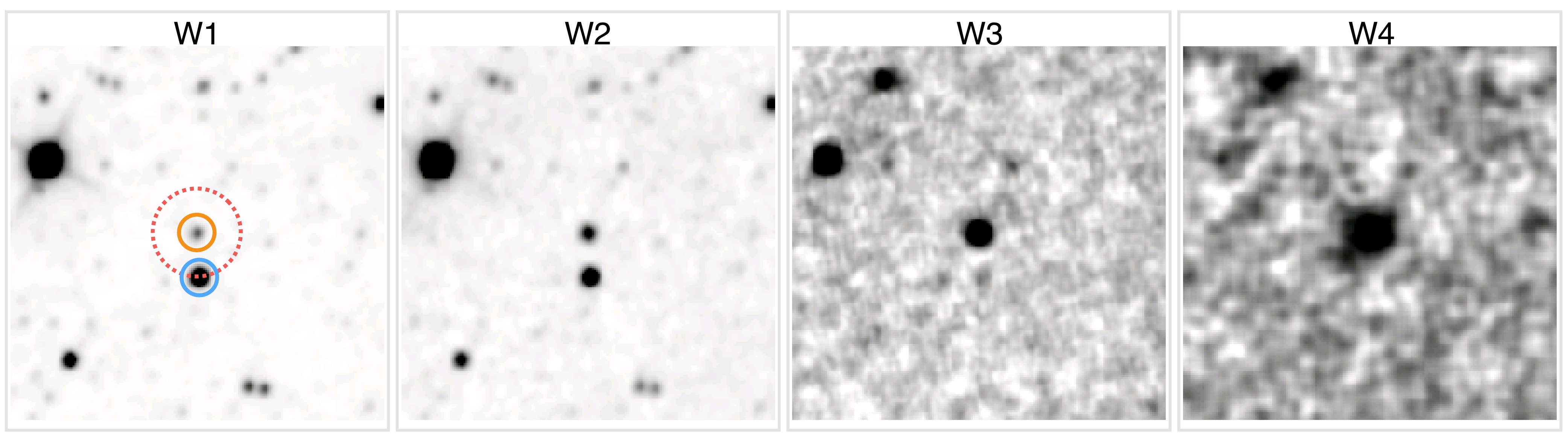}
    \end{center}
    \caption{\wise\ W1--W4 band images of J0916a. North is up, East is left. The size is 5$'\times$5$'$ for each image. The orange circle marks the position of J0916a. One foreground star (blue circle) lies on the edge of \akari\ beam (red dotted circle, with radius of 0.5$'$). }
    \label{fig:image_wise}
\end{figure}

We collected the photometric data for J0916a from SDSS (\textit{u, g, r, i, z}, 3500--9100 \AA, DR7, \cite{Abazajian2009}), \wise\ (W1--W4, 3.4--22 $\mu$m, \cite{Wright2010}), \akari\ (Wide-S, 90~$\mu$m, Yamamura et al., in prep.) and the Very Large Array Faint Images of the Radio Sky at Twenty-cm (VLA FIRST Survey, \cite{Becker1995}). 
\wise\ images of J0916a show one foreground star at the edge of the \akari\ beam in the short wavelength bands (W1 and W2 panels in Figure \ref{fig:image_wise}), but we expect that it does not contaminate the photometry of \akari\, because the star has blue infrared color and the photometry in W4 is dominated by J0916a.

\begin{figure}[!ht]
    \begin{center}
    \includegraphics[width=0.12\textwidth]{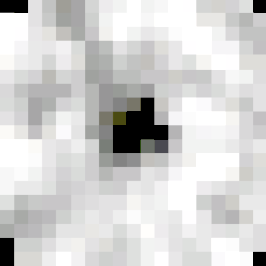}
    \includegraphics[width=0.12\textwidth]{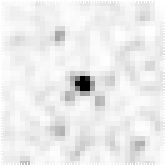}
    \end{center}
    \caption{\akari\ Wide-S 90-$\mu$m and VLA FIRST 21-cm images of J0916a. North is up, East is left. The size is 5$'\times$5$'$ and 1$'\times$1$'$, respectively.}
    \label{fig:image_akari_vla}
\end{figure}


\section{Spectroscopic Properties}
\label{sec:spectral_analyses}

\subsection{Spectral Fitting and Line profiles}
\label{sub:spectral_fitting_and_line_profiles}

\begin{figure*}[!ht]
	\begin{center}
	\includegraphics[trim=0 10pt 0 0, width=\textwidth]{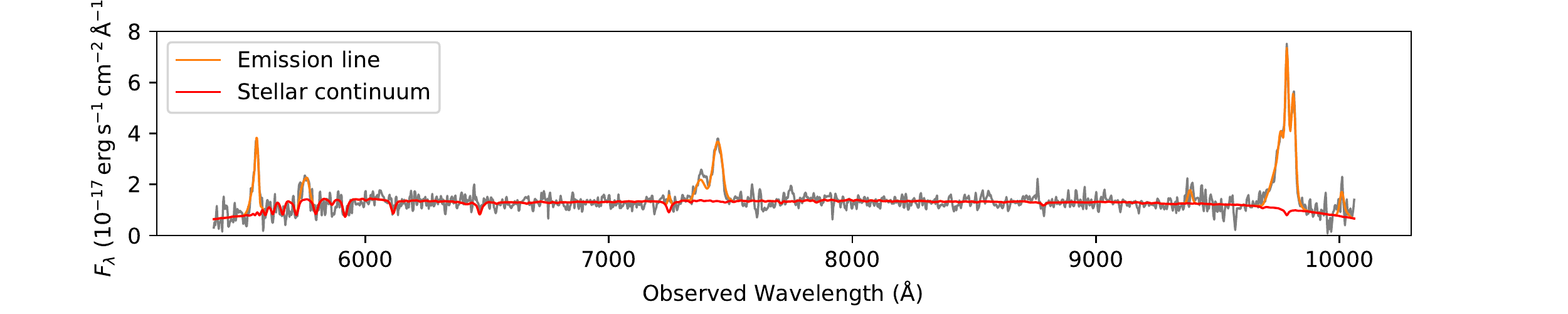}
	\end{center}
	\caption{
	Subaru\,/\,FOCAS spectrum of J0916a in the observed frame. The best-fit stellar continuum and emission lines are shown in red and orange, respectively. 
	}
\label{fig:1dspec_star}
\end{figure*}

\begin{figure}[!ht]
	\begin{center}
	\includegraphics[trim=0 30pt 0 20pt, width=0.5\textwidth]{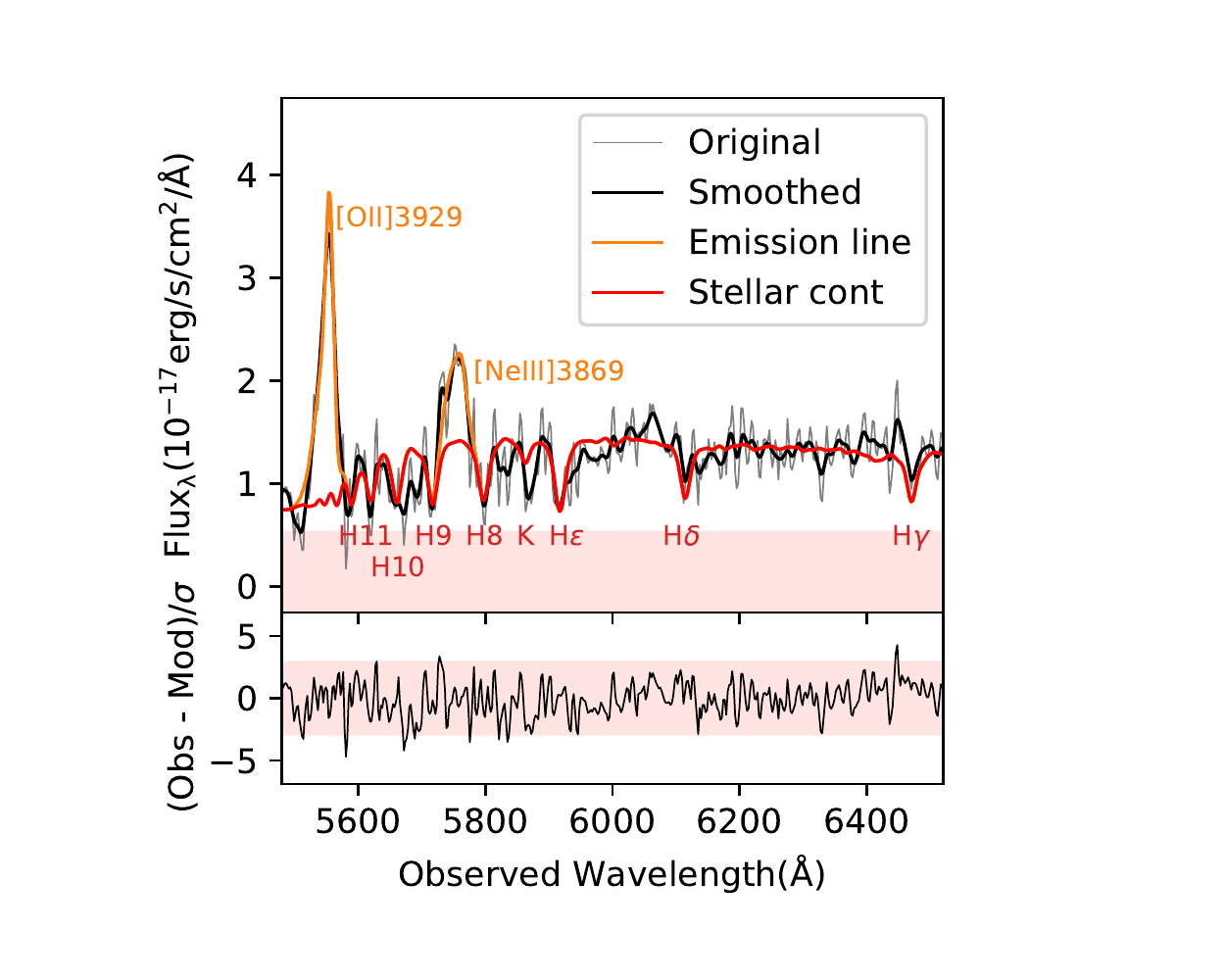}
	\end{center}
	\caption{
	The observed spectrum of J0916a in the Balmer absorption line region (gray solid line) with the best-fit stellar continuum model (red solid line). A series of Balmer absorption features (H$\gamma$--H11) are observed. The bottom panel shows the residual after the subtraction of the stellar continuum model. The pink shading represents the $3\sigma$ level. The noise ($\sigma$) is estimated to be the scatter of the line-free regions. 
	}
	\label{fig:1dspec_star_abs}
\end{figure}

The reduced spectrum of J0916a is shown in Figure \ref{fig:1dspec_star}, which covers the wavelength range from \oii\ doublet to \ha-\nii\ complex. 
In order to examine the emission line properties of J0916a, 
firstly the stellar continuum with absorption lines are subtracted from the galaxy spectrum using the Penalized Pixel-Fitting stellar kinematics extraction code (\texttt{pPXF}), which is a public code for extracting the stellar kinematics and stellar population from absorption line spectra of galaxies \citep{Cappellari2004, Cappellari2012}. 
The code uses the Medium resolution INT Library of Empirical Spectra (\texttt{MILES}) library \citep{Vazdekis2010}, which contains single stellar population synthesis models and covers the full range of the optical spectrum with a resolution of 2.3 \AA\ (FWHM). 
The best luminosity-weighted fitting result is shown as red curve in Figure \ref{fig:1dspec_star}
, and a zoomed-in spectrum covering strong Balmer absorption lines (H$\epsilon$, H8, H9, etc.) is shown in Figure \ref{fig:1dspec_star_abs}. 
The systemic redshift is estimated to be $0.4907\pm0.0001$. 
In order to determine the uncertainty in the systemic redshift, we generate artificial spectra by adding random noise with scatter equal to the standard deviation of the fitting residuals and then perform the same fitting procedure for the artificial spectra using pPXF. The procedures are repeated for 100 times and the scatter is adopted as the uncertainty of the measurement. 
\footnote{
For the uncertainty of the systemic velocity determination, we also consider the difference from the velocity of the narrow \ha\ line. 
We take the velocity of narrow \ha\ emission line in relative to stellar absorption lines as the upperlimit of the uncertainty of the systemic velocity, which is measured as approximately 90 \kms, or 0.0003 in redshift. 
}
The measured velocity dispersion of the stellar component is about 370 \kms\ after correction for the instrumental profile. 
The luminosity-weighted age distribution indicates a main stellar population of 2 Gyr and a starburst population of 70 Myr. 

\begin{figure*}[!ht]
	\begin{center}
	\includegraphics[trim=0 30pt 0 0, width=\textwidth]{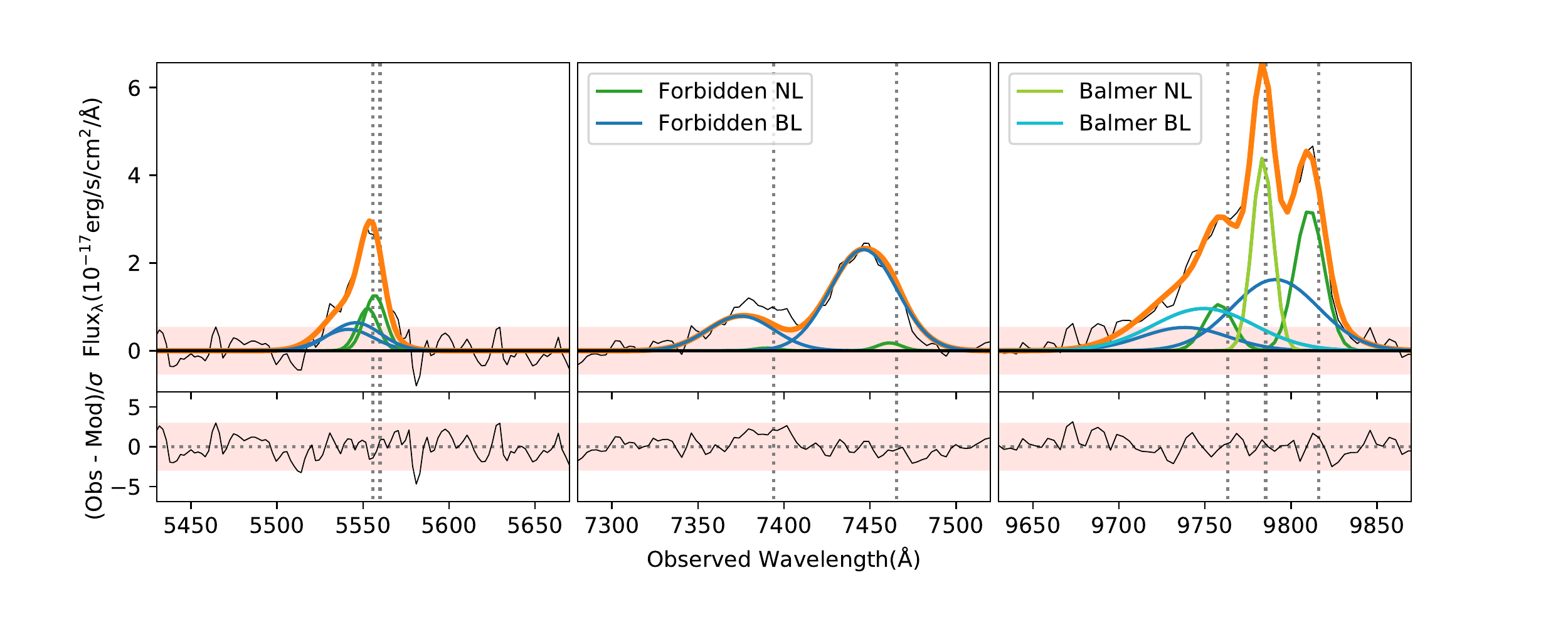}
	\end{center}
	\caption{
	Emission line spectra of \oii\ (left), \oiii\ (middle), and \ha-\nii\ (right). The forbidden lines are fitted with a narrow (green) and a broad (blue) Gaussian components. The narrow and broad Balmer line components are shown in olive and cyan, respectively. The dotted vertical lines denote the systemic position of each emission line. 
	}
	\label{fig:1dspec_lines}
\end{figure*}

The emission line profile fitting is carried out using the Python Spectroscopic Toolkit (\texttt{PySpecKit}) developed by \citet{Ginsburg2011}. 
At first, we examine the profiles of the \oiiilong\ doublet since they are bright and isolated from other emission lines. We tie the central wavelengths and dispersions of the Gaussian profiles for the \oiii\ doublet, and fix the intensity ratio of the two lines to the theoretical value of 1:2.92 \citep{Osterbrock2006}. 
The velocity shift and FWHM of \oiii\ line is $-770$ \kms\ and $1830$ \kms, respectively, if a pair of tied Gaussian profiles is used to fit \oiii\ doublet. The residual shows a blue excess with shift of over $2000$ \kms\ but the excess is within $3\sigma$ level, thus an additional Gaussian profile is not adopted in the following fitting procedure. 
No narrow Gaussian component is required to explain the profile of the \oiii\ doublet. 

We then fit the \oii\ doublet and \ha-\nii\ complex regions. 
The intensity ratio of \niilong\ doublet is fixed to the theoretical value of 1:3.06 \citep{Osterbrock2006}. 
We also fix the intensity ratio of \oiilong\ doublet to 1:1.31. 
The line ratio is calculated with the empirical function of \citet{Sanders2016} by assuming electron temperature of $10^4$ K and electron density of 100 cm$^{-3}$, which are the typical photoionization equilibrium temperature and electron density of \hii\ regions \citep{Sanders2016} and AGN NLR \citep{Netzer1990}. 
We simultaneously fit the forbidden lines, by tying the centers and dispersions of the multiple emission lines for each component\footnote{Note that tying the dispersion here means the assumption that for a certain component every emission line shows the same velocity dispersion after correcting for the instrumental broadening. Thus for each component the width $\sigma$ of the Gaussian profiles is defined as:
\begin{eqnarray*}
    \ \ \ \sigma \equiv \frac{1}{2\sqrt{2\ln2}}\sqrt{ \left( \frac{\Delta v}{c}\lambda_{\rm line} \right) ^2 + \Delta\lambda_{\rm ins}^2 },
\end{eqnarray*} 
\ \ with the same velocity FWHM $\Delta v$ for each line, where the $\lambda_{\rm line}$ and $\Delta\lambda_{\rm ins}$ are the observed central position of the emission lines and instrumental broadening FWHM, respectively. }.
In the case where only a narrow Gaussian profile ($<500$ \kms) is adopted for \oii\ line, a residual excess of 6--7$\sigma$ arises on the blue edge of the \oii\ line, which indicates that another broad component is required to fit this blue wing. Owing to the blending of \ha\ and \nii\ lines, it is hard to determine whether the broad \nii\ lines exist or not. Since IP of \nii\ (14.53 eV) is close to IP of \oii\ (13.62 eV), we assume \nii\ line has the same profile as the \oii\ line and consider the case with broad components of \nii\ doublet (Figure \ref{fig:1dspec_lines}) as a nominal fitting result, hereafter. In order to reduce the degeneracy, the shift and width of broad \oii\ and \nii\ lines are tied to the values measured from the \oiii\ line.

\begin{longtable}{c|ccccc}
\caption{Line properties determined in the best fitting profile (bottom panel of Figure \ref{fig:1dspec_lines}).}\label{tab:spec_fitting}
\hline
\hline
Name & $v_{50}$ (\kms) & FWHM (\kms) & $w_{80}$ (\kms) & Flux ($10^{-16}$\fluxcgs) & $L_{\rm obs}$ ($10^{41}$\lumcgs) \\
\hline
\oiilong\      & $-380\pm90 $ &  $910\pm50$  & $1690\pm50$  &  $7.6\pm0.2$ &  $7.0\pm0.2$ \\
\neiiilong\    & $-670\pm110$ & $1800\pm110$ & $1910\pm70$  &  $3.3\pm0.3$ &  $3.0\pm0.2$ \\
\hb\           &  $-50\pm90 $ &  $310\pm50$  &  $380\pm50$  &  $0.9\pm0.2$ &  $0.9\pm0.2$ \\
\oiiiblong\    & $-770\pm100 $ & $1830\pm70$  & $1940\pm70$  & $11.6\pm0.3$ & $10.7\pm0.3$ \\
\ha\           & $-240\pm90 $ &  $360\pm40$  & $2010\pm140$ & $13.8\pm1.0$ & $12.8\pm0.9$ \\
\niiblong\     & $-380\pm90 $ &  $720\pm50$  & $1680\pm50$  & $17.2\pm0.8$ & $16.0\pm0.7$ \\
\endhead
\hline
\hline
\end{longtable}

In order to describe the multi-component profiles, following Zakamska et al. (2014, 2016) and Perna et al. (2015), 
we use velocity defined from the normalized cumulative distribution: 
\begin{equation}\label{equ:w80}
    F=\int_{-\infty}^{v_F}f(v')\;\mathrm{d}v' \,/\, \int_{-\infty}^{\infty}f(v')\;\mathrm{d}v',
\end{equation}
where $f(v)$ is the flux per velocity unit from the best fit model, $F$ equals to the fraction of flux with velocity $v\le v_F$. 
The velocity shift is defined to be $v_{50}$. 
The width comprising 80 percent of the total flux, $w_{80}\equiv v_{90}-v_{10}$, is equivalent to $2.563\ \sigma$ for a single Gaussian profile. 
We also define the total FWHM as the width at half maximum of the total profile consisting of both of the narrow and broad components. 
The $v_{50}$, FWHM, and $w_{80}$ of each emission line are listed in Table \ref{tab:spec_fitting}. 
The uncertainty for the velocity shifts and widths are estimated using a Monte Carlo method. 
We generate artificial spectra by adding random noise with scatter equal to the standard deviation of the fitting residuals and then perform the same fitting procedure for the artificial spectra. 
The procedures are repeated for 100 times and the $1\sigma$ dispersion is used as the measurement error for each parameter.

Signatures of an extremely strong outflow are indicated in the emission line profiles.
The \oiii\ line shows FWHM of $1830\pm70$ \kms\ with velocity shift of $-770\pm100$ \kms\ in relative to the stellar absorption lines
\footnote{
The uncertanty of the velocity shift is estimated to be $\sqrt{\sigma_{\rm shift,line}^2+\sigma_{\rm system}^2}$, where $\sigma_{\rm shift,line}$ and $\sigma_{\rm system}$ are the scatters for emission lines and the systemic velocity, respectively. 
We take $\sigma_{\rm system}=90$ \kms, which is the deviation between the velocities of stellar absorption lines and narrow \ha\ emission line, and can be considered as the upperlimit of the uncertainty of the systemic velocity. 
}. 
More interestingly, the low-ionization \oii\ doublet also shows large FWHM of $910\pm50$ \kms\ with velocity shift of $-380\pm90$ \kms.  
The velocities of \oiii\ and \oii\ indicate the object shows one of the most powerful outflows observed among ULIRGs\,/\,DOGs (Toba et al. 2017; see discussion of Section \ref{sec:discuss_extreme}) at $z<1$. 
The extremely large width of the low-ionization \oii\ line probably suggests that the strong outflow can extend beyond the nuclear\,/\,bulge region of the galaxy.


\subsection{Long-slit spectroscopy 2D image of the emission lines}
\label{sec:2dspec}

\begin{figure*}
    \begin{center}
    \includegraphics[trim=0 12pt 0 0, width=\textwidth]{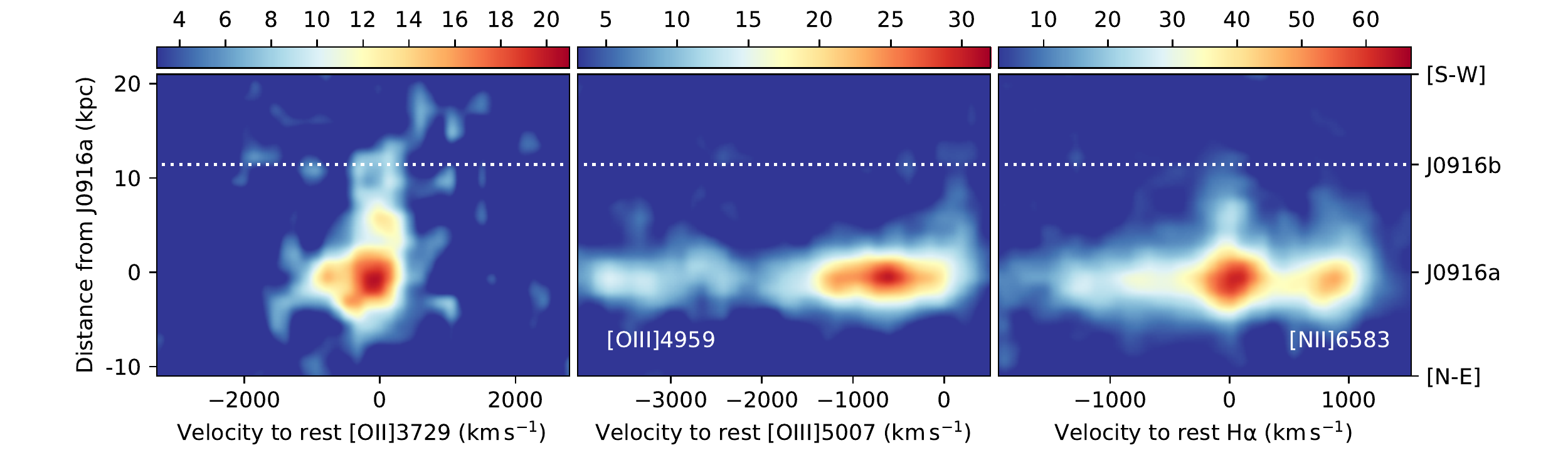}
    \end{center}
    \caption{
    Long-slit spectroscopy 2D image of \oii\ (left), \oiii\ (middle), and \ha-\nii\ (right) emission lines after subtracting the stellar continuum. 
	The color bars represent the S/N ratios and only regions with S/N\,$>3\sigma$ are shown in each panel. The noise ($\sigma$) is estimated to be the scatter of the off-source region in the spectroscopy 2D image. 
    Companion galaxy (J0916b) locates at $y\sim11$ kpc (shown by the horizontal white dotted line).}
    \label{fig:2dspec_orig}
\end{figure*}

In order to examine the spatial distribution of the outflow along the slit direction, we plot long-slit spectroscopy 2D image of the emission lines after subtracting the stellar continuum components. As shown in Figure \ref{fig:image_focas}, one companion galaxy (hereafter J0916b) exists at approximately $2''$ ($\sim11$ kpc) from J0916a. Therefore the stellar contribution from both of the two galaxies are considered. 
We extract the integrated stellar spectrum of J0916b using the same method as for the J0916a. 
It is hard to fit the stellar continuum for each line of the data due to the low S\,/\,N ratio at the outer regions. 
Thus, we assume the inner and outer regions of each galaxy have the same stellar continuum, and apply a Gaussian profile to normalize the spatial distribution. 
The Gaussian profile is determined using the integrated flux distribution in line-free regions. 
The emission line components can be obtained after subtracting the stellar components of these two galaxies. Three zoom-in views around \oii, \oiii, and \ha-\nii\ emission lines are shown in Figure \ref{fig:2dspec_orig}. 
Only regions with S/N\,$>3\sigma$ are shown in each panel. The noise ($\sigma$) is estimated to be the scatter of the off-source region in the spectroscopy 2D image. 

J0916b lies at the same redshift ($z=0.49$) as J0916a. 
Among the three zoom-in views in Figure \ref{fig:2dspec_orig}, J0916b only has relatively luminous \oii\ line, which is connected to the extended \oii\ component of J0916a. 
The \oii\ `bridge' component possibly indicates a direct interaction between the two galaxies, 
which is consistent with the tidal optical morphology (Figure \ref{fig:image_focas}), 
and also consistent with the evolutionary scenario that ULIRGs are thought to be the descendants of galaxy mergers. 
On the other hand, the outflowing gas of J0916a shows extended structure (\oii\ and \oiii\ panels in Figure \ref{fig:2dspec_orig}), which supports a possible evidence of the galaxy-wide outflow. 
We discuss the velocity structure and ionization source further in Section \ref{sec:discuss_extreme}. 


\section{SED Analyses with multi-band photometric data} 
\label{sec:sed_analyses}

\subsection{Building templates} 
\label{sub:models}

The offset between J0916a and J0916b ($\sim$ 2$''$) is smaller than the spatial resolution of \wise\ (6.1--12.0$''$) and \akari\ (1.0--1.5$'$). Since J0916b is fainter in the optical and has only weak emission lines, which suggests no strong AGN or star formation activity, we assume that all of the observed MIR and FIR fluxes are associated with J0916a. 

We perform a SED fitting for J0916a using the Code Investigating GALaxy Emission (\texttt{CIGALE}, \cite{Noll2009, Serra2011}) of version 0.12. 
Stellar continuum in the UV-optical as well as dust emission in the IR band are modeled in \texttt{CIGALE}, assuming the energy balance between absorption and re-emission by interstellar dust. 
AGN component is also modeled with the energy balance between absorption and re-emission by the dusty torus. 
The total IR luminosity between 1 $\mu$m and 1000 $\mu$m, $L_{\rm IR}$, is defined as the sum of the emissions from dust heated by stars and AGN. 
Several parameters need to be set to generate templates. 
The observed photometric data points are then fitted with the templates. 
The best fit template is selected with $\chi^2$ minimization in the \texttt{CIGALE} code. 

In order to model the stellar continuum, the star formation history (SFH) needs to be assumed. 
We adopt two stellar components, i.e., an old stellar population (OSP) with an exponentially declining SFR and a young stellar population (YSP) with a constant SFR \citep[2014]{Buat2011}. 
In order to avoid too many parameters in the SED fitting, only three parameters, i.e., the ages of OSP and YSP as well as the mass fraction of YSP ($f_{\rm burst}$), are used as free parameters to reproduce SFH (see Table \ref{tab:sed_input}). The values of the age parameters are selected based on the estimation from luminosity-weighted fitting result of stellar continuum (Figure \ref{fig:1dspec_star}). The e-folding time ($\tau$) of OSP is fixed to be 500 Myr, assuming a typical quenching timescale for local galaxies \citep{Lian2016}. 
The single stellar population model of \citet{Bruzual2003} with the initial mass function of \citet{Salpeter1955} is adopted. Metallicity is fixed to be the solar metallicity of 0.02. 

\citet{Calzetti2000} dust attenuation law and \citet{Dale2014} dust emission model are adopted in the SED fitting. 
In the dust attenuation model, a reduction factor is applied to the color excess of the old stellar population. 
The reduction factor is defined as 
$f_{\rm att} \equiv E(B-V)_{\rm OSP}/E(B-V)_{\rm YSP} = A_{\rm V}^{\rm ISM}/A_{\rm V}^{\rm BC}$, 
under the assumption that the young stars are embedded in dense molecular birth clouds (BC) and old stars are in diffuse ISM \citep{Calzetti2000, Charlot2000, daCunha2010}. 
In order to reduce the number of free parameters in the SED fitting, we fix $f_{\rm att}=0.5$, following \citet{Buat2012} and \citet{LoFaro2017}. 
In the dust emission model of \citet{Dale2014}, the dust is assumed to be exposed to a distribution of starlight intensities described by a power-law function:
$dM_{\rm dust}/dU=U^{-\alpha_{\rm SF}}$.
The exponent $\alpha_{\rm SF}$ determines the relative contributions of different SED templates, 
which are comprised of calculated emission from small and large dust grains, as well as Polycyclic Aromatic Hydrocarbon (PAH) whose spectrum is derived from the average MIR spectrum of star-forming galaxies observed by the Infrared Space Observatory (\textit{ISO}) satellite. 
A lower $\alpha_{\rm SF}$ indicates stronger heating intensities with the FIR emission peak at shorter wavelengths, and corresponds to more active star-forming activities in a galaxy \citep{Dale2014}. 
Due to the lack of detections in the sub-millimeter bands, it is hard to constrain the FIR emission peak for J0916a. Therefore we only test three values of $\alpha_{\rm SF}$ based on the results in the literature, i.e., 
$\alpha_{\rm SF}\sim1.1$, which corresponds to a SED peaks at around 60\ $\mu$m in the rest frame with a translated grey-body temperature of 40 K, from an \iras\ 60-$\mu$m selected `warm' ULIRG sample at $z=0.1$ \citep{Symeonidis2011}; 
$\alpha_{\rm SF}\sim1.6$ for a SED with a peak at 90--100 $\mu$m and temperature of about 30 K, from the $Spitzer$\,/\,MIPS 70-$\mu$m selected `cold' ULIRG sample at $0.1<z<1.2$ \citep{Symeonidis2011} 
and the 90-$\mu$m selected ULIRGs in the \akari\ Deep Field-South (ADF-S) survey at $0.4<z<1.2$ \citep{Malek2017}; 
and $\alpha_{\rm SF}\sim2.1$ for SEDs of normal galaxies in ADF-S field \citep{Malek2017}. 

The \citet{Fritz2006} model is adopted to generate AGN emission templates. The model consists of three different components: the emission directly from the central source, the scattered component, and the thermal radiation of a dusty torus heated by the central source. 
Six parameters were used by \citet{Fritz2006} to describe the structure and geometry of the dusty torus and to calculate the radiation transfer. 
However, in order to reduce the number of free parameters in the SED fitting with limited photometric data points, 
we fix the radial ($\beta=-0.5$) and angular ($\gamma=0.0$) dust distribution, the opening angle of the torus ($\theta=100.0^{\circ}$) as well as the ratio of the radii ($R_{\rm max}/R_{\rm min}=60$) following \citet{Malek2017}. 
Only two extreme values of the viewing angle of the torus $\psi$, 0.001$^{\circ}$ and 89.990$^{\circ}$, are considered, which corresponds to type-2 and type-1 AGNs, respectively. 
Silicate optical depth $\tau_{9.7}$ and AGN fraction $f_{\rm IR}^{\rm \scriptscriptstyle AGN}$, which is defined as the ratio of AGN IR luminosity to the total IR luminosity of the galaxy, are chosen as free parameters in the SED fitting procedure with ranges of 1.0--6.0 and 0.01--1.0, respectively. 
Although AGN IR emission is also considered in the \citet{Dale2014} model, we exclude this component because the templates are mainly from type-1 AGNs and do not have the absorption features in the MIR band which are observed in type-2 AGNs. 

Since J0916a was also detected at 1.4 GHz in the VLA FIRST survey, we consider the radio non-thermal component in the SED fitting. 
Two parameters, FIR\,/\,radio correlation coefficient ($q_{\rm FIR}$) and slope of the power-law synchrotron emission ($\alpha_{\rm rad}$) are fixed to be 
2.58 and 0.70, respectively, based on the model of \citet{Colina1992}, in which the non-thermal radio emission from radio supernovae and supernova remnants are the main contributor to the radio luminosity in the ULIRGs. A fixed $q_{\rm FIR}$ results in the constraint on the ratio between the FIR and radio luminosities of J0916a. 
We also test other values of $q_{\rm FIR}$ from the literature, e.g., 2.1 for sub-millimeter galaxies (SMGs) and ULIRGs at $1<z<3$ \citep{Kovacs2006}, 2.3 for radio-identified, flux-limited ($S_{\rm60\mu m}\ge2$ Jy) \iras\ sources \citep{Yun2001}, and 2.8 for the local spirals and starburst galaxies \citep{Condon1991}. 
If we fit the SED with changing $q_{\rm FIR}$ of 2.1, 2.3, 2.58, and 2.8, 
the reduced $\chi^2$ for the best fit cases are 11.4, 6.3, 0.8, and 3.7, respectively, 
with the estimated $L_{\rm IR}$ from interstellar dust
varying between $2.6\times10^{12}$ and $6.4\times10^{12}$ \lumcgs. 
The value of $q_{\rm FIR}$ is hard to be constrained with the current limited data points, especially in the FIR wavelength range. 
Therefore we select $q_{\rm FIR}=2.58$, since it results in the smallest $\chi^2$. 
The values and ranges of the free parameters are summarized in Table \ref{tab:sed_input}.

In order to estimate the uncertainties of the output parameters, we perform the same fitting procedure to the simulated mock SED. 
The mock SEDs are generated by adding random values with the same scatters as the measurement errors to the observed photometric data points in each band. We repeat the fitting procedure for 100 mock SEDs and use resulting standard deviations as the $1\sigma$ uncertainties of the output parameters. 

\begin{table}[!ht]
\caption{Parameters of the \texttt{CIGALE} SED fitting.}\label{tab:sed_input}
\begin{tabular}{ll}
\hline
\hline
\multicolumn{2}{c}{Star formation history}  \\ 
OSP e-folding time $\tau$ (Myr) & 500 \\
Age of OSP $t_{\rm main}$ (Myr) & 100, 200, 400, 600 \\
Age of YSP $t_{\rm burst}$ (Myr) & 10, 20, 40, 80 \\
Mass fraction of YSP $f_{\rm burst}$ & 0.01, 0.03, 0.05, 0.1, 0.3, 0.5 \\
\hline
\multicolumn{2}{c}{Single stellar population \citep{Bruzual2003}} \\
Initial mass function & Salpeter (1955) \\
Metallicity & 0.02 \\
\hline
\multicolumn{2}{c}{Dust attenuation \citep{Calzetti2000}} \\
E(B-V) of YSP & 0.2--2.0, per 0.2 \\
Reduction factor $f_{\rm att}$  & 0.5 \\
\hline
\multicolumn{2}{c}{Dust re-emission \citep{Dale2014}} \\
Dust heating slope $\alpha_{\rm SF}$  & 1.125, 1.625, 2.125 \\
\hline
\multicolumn{2}{c}{AGN emission \citep{Fritz2006}} \\
$R_{\rm max}/R_{\rm min}$ of dusty torus & 60 \\
Optical depth at 9.7 $\mu$m $\tau_{9.7}$ & 1.0, 3.0, 6.0 \\
Radial dust distribution $\beta$ & $-$0.5 \\
Angular dust distribution $\gamma$  & 0.0 \\
Opening angle of torus $\theta$ & 100.0$^{\circ}$ \\
Angle of viewing axis $\psi$ & 0.001$^{\circ}$, 89.990$^{\circ}$ \\
$f^{\rm \scriptscriptstyle AGN}_{\rm IR}=L_{\rm IR}^{\rm \scriptscriptstyle AGN}/L_{\rm IR}^{\rm tot}$ & 0.01--0.09, per 0.01\\
 & 0.1--1.0, per 0.05 \\
\hline
\multicolumn{2}{c}{Radio non-thermal emission \citep{Colina1992}} \\
FIR\,/\,radio coefficient $q_{\rm FIR}$ & 2.58   \\
Radio power-law slope $\alpha_{\rm rad}$ & 0.7 \\
\hline
\hline
\end{tabular}
\end{table}

\begin{table}[!ht]
\caption{Estimation of galaxy properties from the best fit SED$^{\star}$.}
\label{tab:sed_output}
\begin{tabular}{ll}
\hline
\hline
SFR estimated from $L_{\rm IR}^{\rm dust}$ & $990\pm44$ \sfrunit \\
Stellar mass $M_{\star}$ & $9.46\pm1.69\times10^{10}$ \ms \\
Stellar un\_att luminosity $L_{\rm un\_att}^{\rm star}$ & $6.86\pm0.31\times10^{12}$ \ls \\
\hline
$E(B-V)$ of YSP   & $1.49\pm0.10$ \\
\hline
SF IR luminosity $L_{\rm IR}^{\rm dust}$  & $5.75\pm0.26\times10^{12}$ \ls \\
Total IR luminosity $L_{\rm IR}^{\rm tot}$  & $6.13\pm0.25\times10^{12}$ \ls \\
\hline
AGN bolometric luminosity $L_{\rm bol}^{\rm \scriptscriptstyle AGN}$  & $4.65\pm0.39\times10^{11}$ \ls \\
$f^{\rm \scriptscriptstyle AGN}_{\rm IR}=L_{\rm IR}^{\rm \scriptscriptstyle AGN}/( L_{\rm IR}^{\rm \scriptscriptstyle AGN} + L_{\rm IR}^{\rm dust} )$    & $6.27\pm0.66\,\%$ \\
$f^{\rm \scriptscriptstyle AGN}_{\rm bol}=L_{\rm bol}^{\rm \scriptscriptstyle AGN}/( L_{\rm bol}^{\rm \scriptscriptstyle AGN} + L_{\rm un\_att}^{\rm star} )$ & $6.34\pm0.65\,\%$ \\
\hline
\hline
\multicolumn{2}{l}{$^{\star}$in the case of $\alpha_{\rm SF}=1.125$ and $q_{\rm FIR}=2.60$} \\
\end{tabular}
\end{table}

\begin{figure}[!ht]
    \begin{center}
    \includegraphics[trim=0 6pt 0 22pt, clip, width=0.5\textwidth]{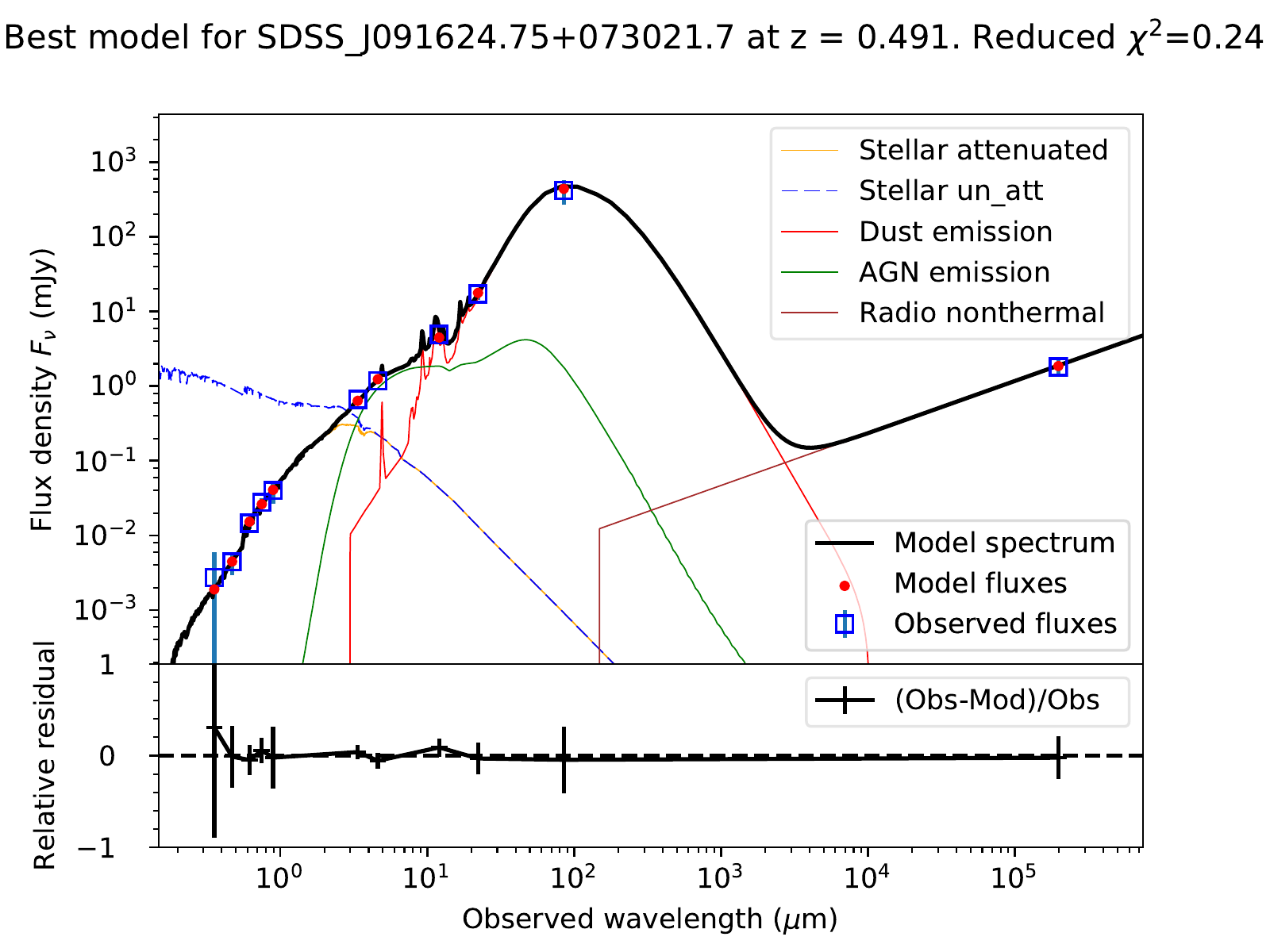}
    \end{center}
    \vspace{-4pt}
    \caption{Best fit SED result of J0916a. 
    The observed data points are shown with blue squares, while the model SED is shown with a black thick curve. 
    The intrinsic and attenuated stellar radiation components are marked with blue dashed and orange solid curves, respectively, while the thermal re-emission by interstellar dust is shown wuth a red solid curve. 
    The AGN component and radio synchrotron radiation are denoted with green and brown solid curves, respectively. 
    }
    \label{fig:sed_best}
\end{figure}

\begin{figure}[!ht]
    \begin{center}
    \includegraphics[trim=0 6pt 0 22pt, clip, width=0.5\textwidth]{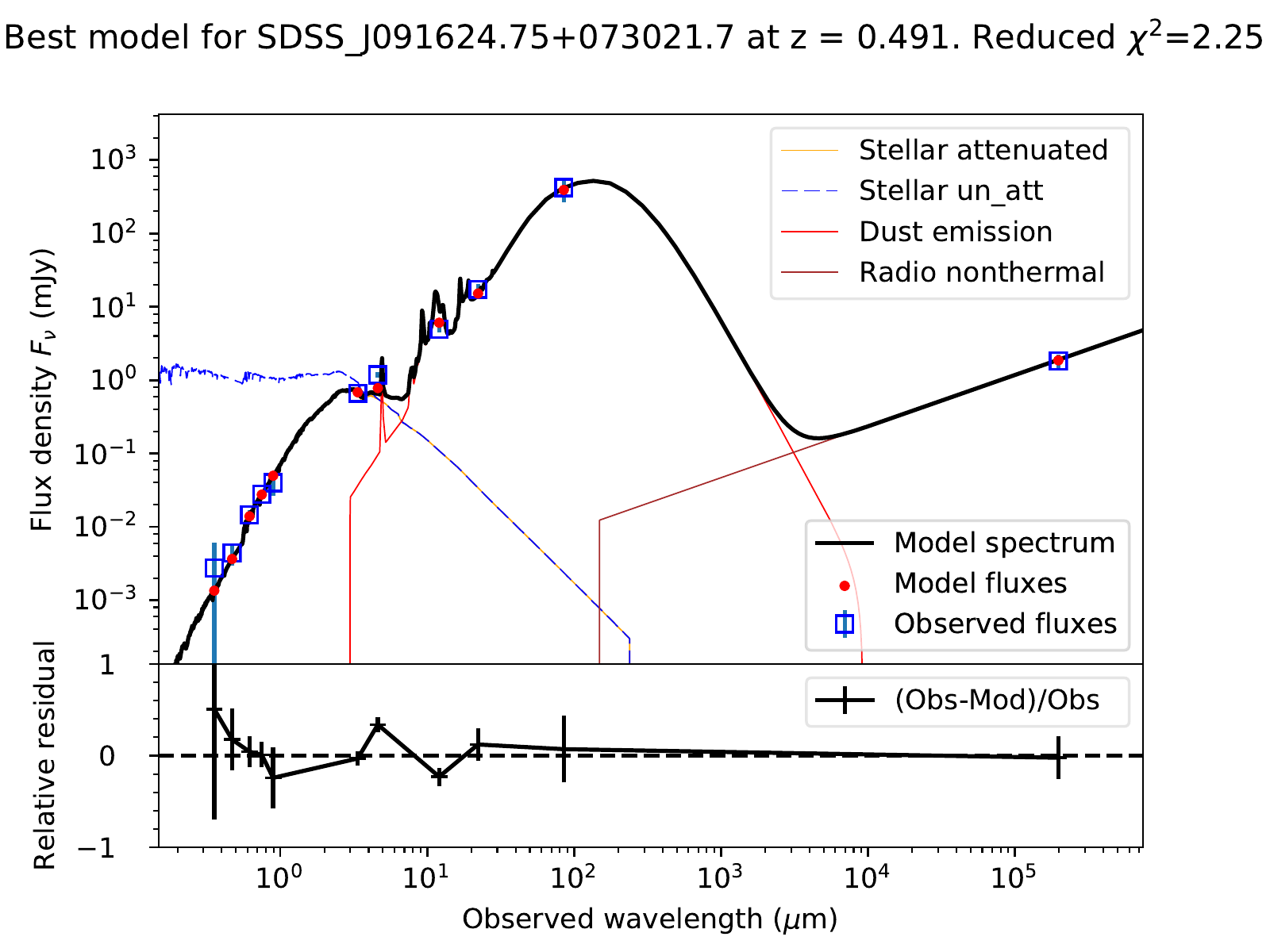}
    \end{center}
    \vspace{-4pt}
    \caption{SED fitting result of J0916a with AGN contribution fixed to 0\,\%. 
    The legends are the same as Figure \ref{fig:sed_best}. Compared to $\chi^2=0.79$ of the best fit case, the fitting quality without AGN component decreases to $\chi^2=7.50$.}
    \label{fig:sed_comp}
\end{figure}

\subsection{The best fit properties from the SED fitting}
\label{sec:sed_output}

The best fit parameters and SED model of J0916a are shown in Table \ref{tab:sed_output} and Figure \ref{fig:sed_best}, respectively. 
The resulting reduced $\chi^2=0.79$ suggests that the model with relatively large number of free parameters (8) does not overfit the observation with limited number of data points (11). 

The total IR luminosity between 1 $\mu$m and 1000 $\mu$m in the rest frame reaches \lumir\ $= 6.13\pm0.25\times10^{12}$ \ls.
94\,\% of the IR luminosity comes from the re-emission of the interstellar dust heated by star formation, and the component dominates the FIR peak. 
Using the empirical relationship of SFR (\sfrunit) $=4.5\times10^{-44}\ L_{\rm IR}^{\rm dust}$ (\lumcgs) \citep{Kennicutt1998}, the SFR is estimated to be 990 \sfrunit. 
With the estimation of stellar mass of $9.46\times10^{10}$ \ms, J0916a lies above the star-forming main sequence at $0.0<z<0.7$ \citep{Tasca2015}, and the deviation indicates strong starburst activity in this object. 
The reduced $\chi^2$ for the cases where $\alpha_{\rm SF}$ equals to 1.125, 1.625, and 2.125 are 0.79, 4.09, and 22.87, respectively, which suggests that J0916a is similar with the `warm' ULIRG samples with relatively large grey-body temperature ($\sim40$ K, \cite{Symeonidis2011}). 

In \texttt{CIGALE} the contribution of AGN component (green solid curve in Figure \ref{fig:sed_best}) is defined as $f^{\rm \scriptscriptstyle AGN}_{\rm IR}\equiv L_{\rm IR}^{\rm \scriptscriptstyle AGN}/L_{\rm IR}^{\rm tot}$, where $L_{\rm IR}^{\rm \scriptscriptstyle AGN}=L_{\rm thermal}^{\rm \scriptscriptstyle AGN}$ is the thermal emission of dusty torus, and the fraction is estimated to be $6.27\pm0.66\,\%$. 
We can also estimate the AGN contribution using the bolometric luminosity of the AGN, which is calculated as the sum of the luminosities of the three components in the model of \citet{Fritz2006}:
\begin{equation}
    L_{\rm bol}^{\rm \scriptscriptstyle AGN}=L_{\rm central}^{\rm \scriptscriptstyle AGN}+L_{\rm scattering}^{\rm \scriptscriptstyle AGN}+L_{\rm thermal}^{\rm \scriptscriptstyle AGN},
\end{equation} 
where $L_{\rm central}^{\rm \scriptscriptstyle AGN}$ is the isotropic luminosity of the central energy source integrated in the range of 0.001--20 $\mu$m, $L_{\rm scattering}^{\rm \scriptscriptstyle AGN}$ and $L_{\rm thermal}^{\rm \scriptscriptstyle AGN}$ are luminosities of scattering and thermal emission from dusty torus, respectively, integrated in the range of 1--1000 $\mu$m. 
In the best fit SED of J0916a, the AGN emission only dominates the MIR band and does not extend to the range shorter than 1 $\mu$m, indicating that the radiation directly from the central accretion disc is obscured in the UV-optical range. The obscuration of AGN is consistent with the extended optical morphology (see Figure \ref{fig:image_focas}). 
The bolometric luminosity of the host galaxy can be estimated as:
\begin{equation}
    L_{\rm bol}^{\rm host}=L_{\rm attenuated}^{\rm star}+L_{\rm IR}^{\rm dust}=L_{\rm un\_attenuated}^{\rm star}.
\end{equation} 
Therefore, we obtain $f^{\rm \scriptscriptstyle AGN}_{\rm bol}\equiv L_{\rm bol}^{\rm \scriptscriptstyle AGN}/(L_{\rm bol}^{\rm \scriptscriptstyle AGN}+L_{\rm bol}^{\rm host})=$ \mbox{$6.34\pm0.65\,\%$}, which is consistent with $f^{\rm \scriptscriptstyle AGN}_{\rm IR}$. 
The bolometric luminosity of obscured AGN is assumed to be the sum of total IR luminosity and X-ray luminosity in several works
\citep{Vasudevan2010, Lusso2011}. 
We do not have the X-ray constraint on the AGN activity of J0916a. 
However, the intrinsic 2--10 keV to AGN bolometric luminosity is 0.03\,\%--0.81\,\% in the AGN hosted ULIRGs \citep{Teng2015} or 4\,\%--8\,\% in obscured AGNs \citep{Lusso2011}. Thus the lack of X-ray observation will not affect the estimation of $L_{\rm bol}^{\rm \scriptscriptstyle AGN}$ of J0916a. 

The resulting contribution of AGN is significantly smaller than the average AGN fraction observed among ULIRGs (35\,\%--40\,\%, \cite{Veilleux2009}; 40\,\%--50\,\%, \cite{Goto2010}; 20\,\%, \cite{Ichikawa2014}; 19\,\%, \cite{Malek2017}) and even smaller than that of LIRGs (12\,\%, \cite{Alonso2012}; 10\,\%, \cite{Buat2015}; 12\,\%, \cite{Malek2017}). 
In order to test the necessity of the AGN component in the SED fitting, 
we perform the same fitting procedure with AGN contribution fixed to be 0\,\%, and the result is shown in Figure \ref{fig:sed_comp}. 
The reduced $\chi^2$ significantly increase to 7.50, which suggests that the AGN component is necessary to reproduce the MIR photometry.


\section{Discussion}
\label{sec:discuss}

\subsection{Outflow structure and ionization source}
\label{sec:discuss_extreme}

\begin{figure}[!ht]
    \begin{center}
    \hspace{-11pt}
    \includegraphics[trim=0 25pt 0 0, width=0.5\textwidth]{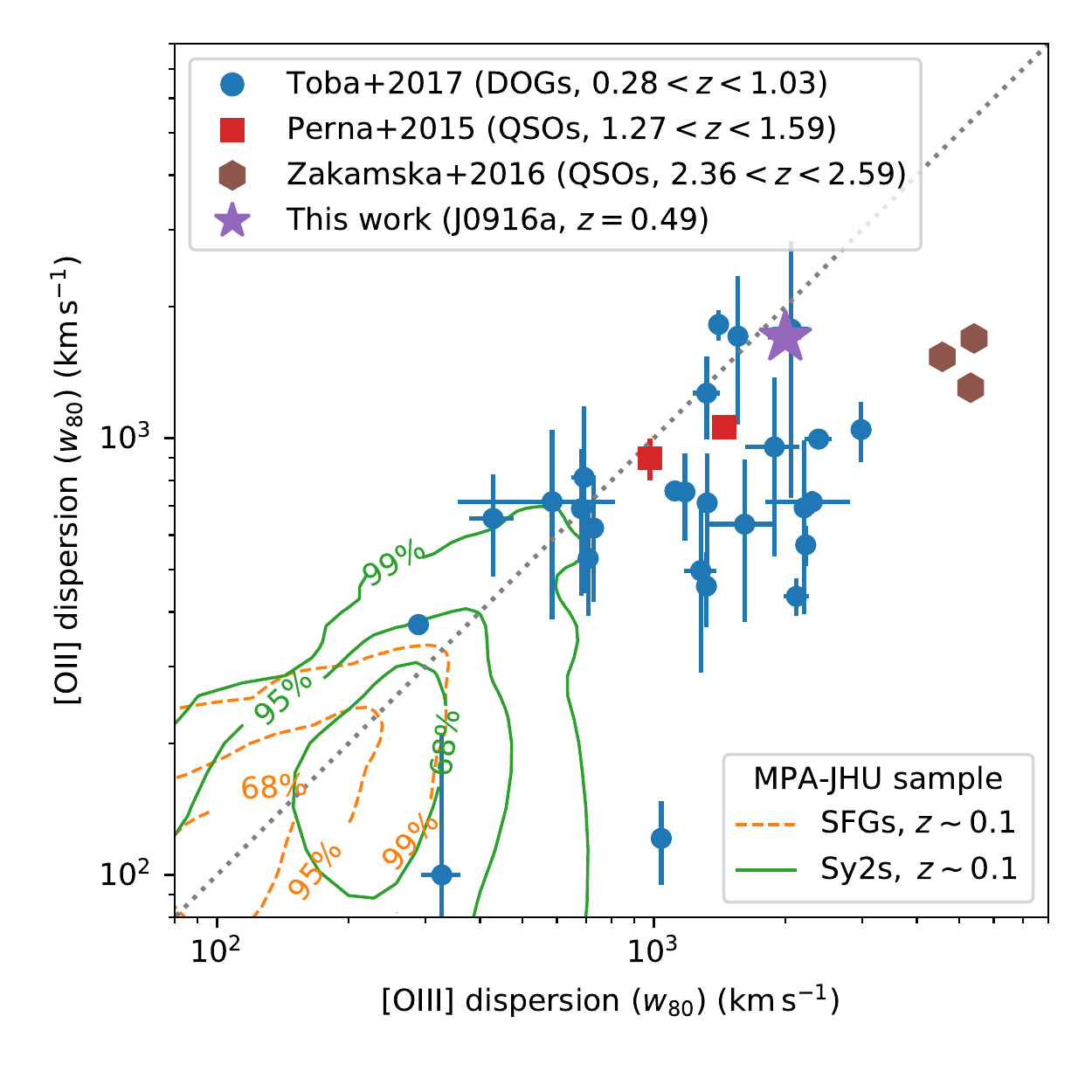}
    \end{center}
    \caption{Velocity dispersions ($w80$) of \oiii\ and \oii\ of J0916a, compared with those of the local star-forming galaxies (SFGs) and Seyfert 2 galaxies (Sy2s) at $z\sim0.1$ (orange dashed and green solid contours, respectively, MPA-JHU), IR-bright DOGs at $0.3<z<1.0$ (blue circles, Toba et al. 2017), and high redshift obscured QSOs at $z\sim1.5$ (red squares, Perna et al. 2015) and at $z\sim2.5$ (brown hexagons, Zakamska et al. 2016). Note that since Zakamska et al. (2016) did not present the results of \oii\ line width, here we use the width of \nii\ instead, because \nii\ and \oii\ have similar ionization potential (Zakamska et al. 2014). The dotted line shows 1:1 relation. 
    }
    \label{fig:comp_v_out}
\end{figure}

It is commonly found that highly ionized gas, e.g., \oiii\ and \neiii, tends to show strong outflows. 
J0916a shows extreme outflow not only in the \oiii\ emission line, but also in the \oii\ emission line.  
In order to compare the broad line widths of both \oiii\ and \oii\ emission lines with galaxies in the local universe, 
we collect the samples in the literature with both \oiii\ and \oii\ measurements. 
We examine the line widths of 1.8 million SDSS DR7 galaxies at $z\sim0.1$, whose absorption and emission profiles are measured by the MPA-JHU group\footnote{https://wwwmpa.mpa-garching.mpg.de/SDSS/DR7/}. 
Only the objects with $S/N>7$ for both \oiii\ and \oii\ emission lines are selected, and their velocity distributions ($w80$) are shown in Figure \ref{fig:comp_v_out} with contours. 
We separate the sample into star-forming galaxies (SFGs) and Seyfert 2 galaxies (Sy2s) 
following the classification of \citet{Brinchmann2004} with the ``Baldwin, Phillips \& Terlevich'' (BPT) diagram \citep{Baldwin1981, Osterbrock2006}. 
The distributions of SFGs and Sy2s show peak at 100 \kms\ and 200 \kms, respectively. 
J0916a lies far beyond the velocity range of both SFG and Sy2 samples in the local universe. 

Recently several works have been presented on strong outflows detected in dust obscured quasars\,/\,galaxies at high redshifts. 
\citet{Perna2015} and \citet{Zakamska2016} reported the discovery of very broad (FWHM $=1500\textrm{--}3000$ \kms) and strongly blueshifted \oiii\ emission line in the spectra of obscured quasars at $z\sim2$. 
\citet{Toba2017} analyzed the ionized gas properties for 36 IR-bright DOGs at $0.3<z<1.0$, of which 25 objects are classified as ULIRGs/HyLIRGs, and found that the IR-bright DOGs show relatively strong outflows compared to Sy2s at $z<0.3$. 
The dispersions of \oiii\ and \oii\ emission lines of these objects are also shown in Figure \ref{fig:comp_v_out}. 
In order to compare with the velocity dispersion measured in the same method, 
we obtain SDSS spectra of the DOG sample in \citet{Toba2017} and re-preform spectral fitting for the DOGs with significant detections of both \oiii\ and \oii\ lines to measure the $w80$ dispersions. 
The powerful outflow of J0916a characterized by the broad ionized oxygen lines is among the fastest ones compared with the ULIRGs/DOGs at intermediate redshifts, and even comparable to the luminous high-redshift QSOs. 

\begin{figure*}[!ht]
	\begin{center}
	\hspace{-20pt}
	\includegraphics[trim=0 16pt 0 0, width=0.333\linewidth]{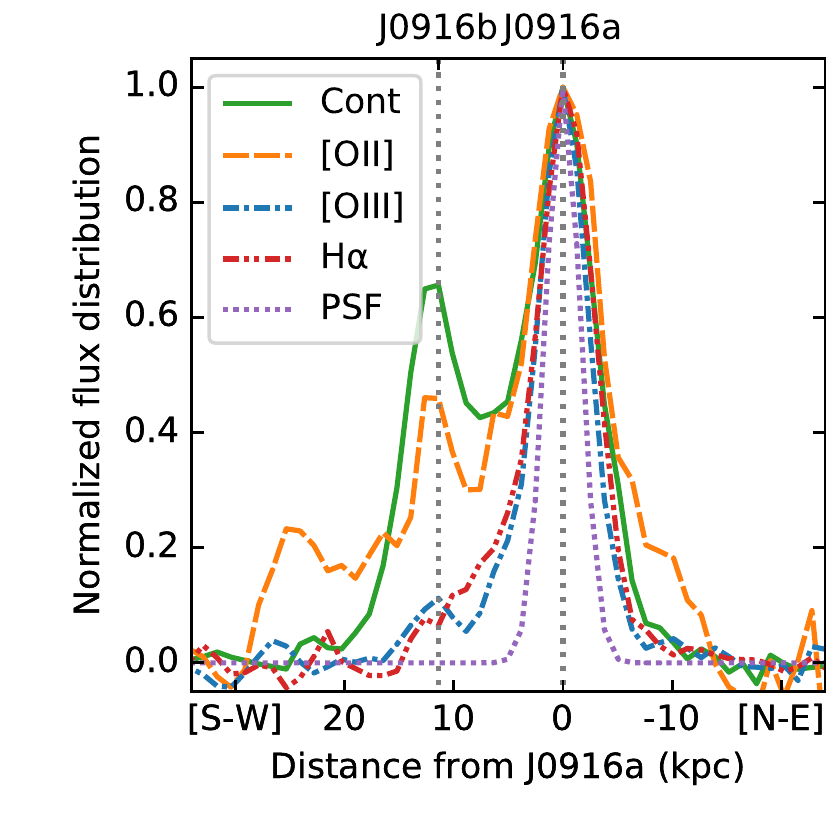}
	\includegraphics[trim=0 16pt 0 0, width=0.333\linewidth]{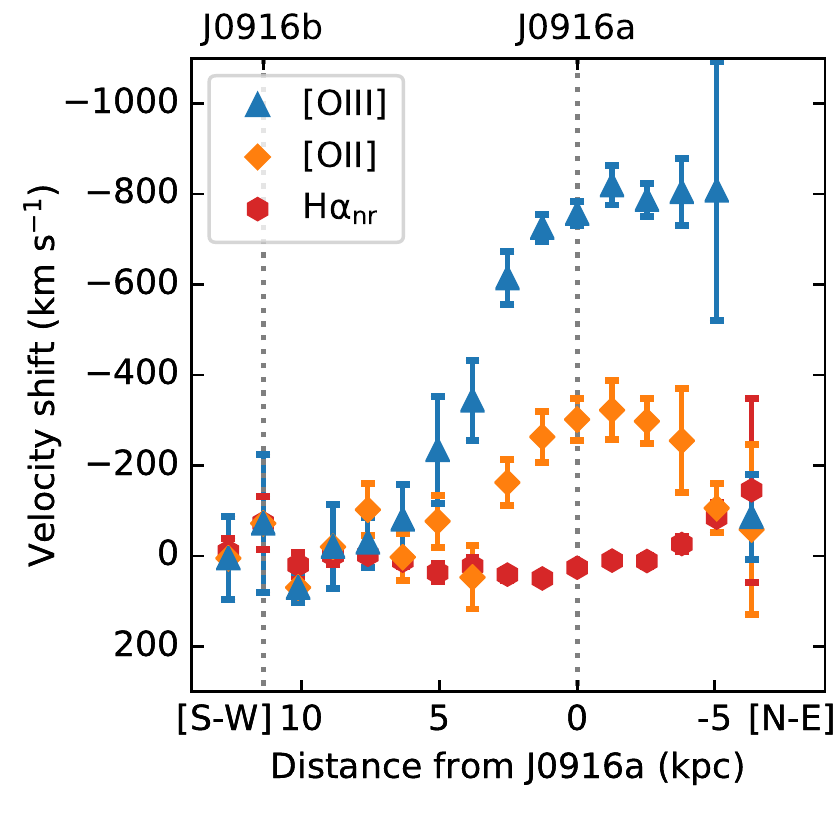}
	\includegraphics[trim=0 16pt 0 0, width=0.333\linewidth]{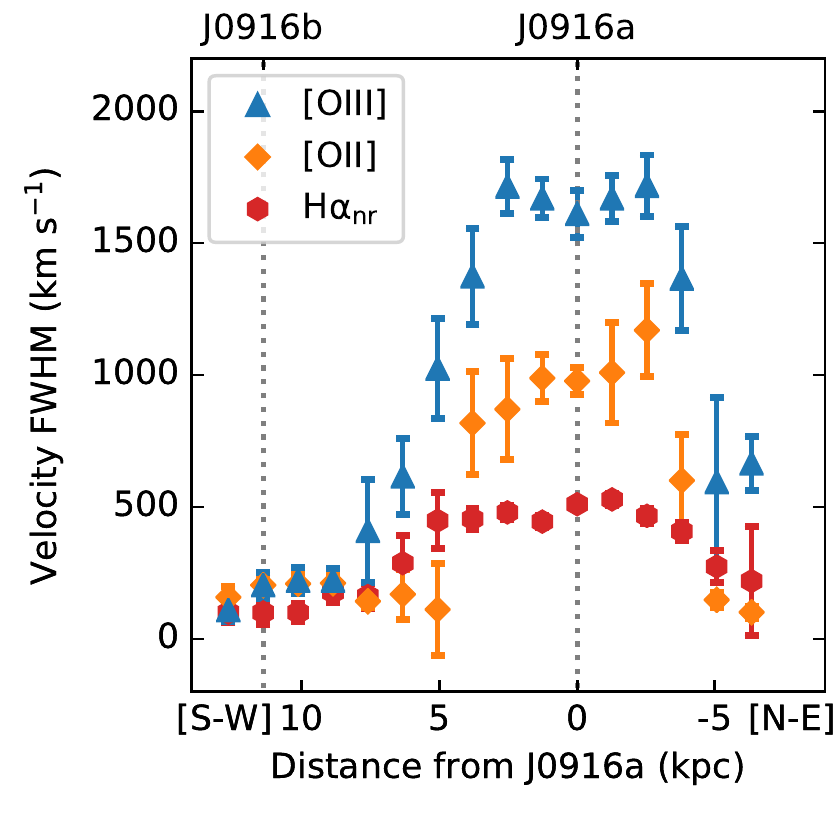}
	\end{center}
    \caption{
    \textbf{Left}: Normalized flux distributions along slit direction for \oii\ (orange dashed), \oiii\ (blue dash-dotted), \ha\ (red dash-dot-dotted) emission lines, and R-band stellar continuum (green solid). The normalized flux of the standard star, Feige34, is also plotted to show the PSF size (violet dotted). 
    \textbf{Middle}: Velocity shift in relative to stellar absorptions of \oiii\ (blue triangles), \oii\ (orange diamonds), and \ha\ (narrow component, red hexagons) emission lines. 
    \textbf{Right}: Velocity dispersion (FWHM) of \oiii\ (blue triangles), \oii\ (orange diamonds), and \ha\ (narrow component, red hexagons) emission lines. 
    In all of the three panels, the vertical dotted lines denote the positions of J0916a and J0916b. 
    }
    \label{fig:2dspec_dist_orig}
\end{figure*}

In order to examine the spatial extent of the ionized outflow, we perform spectral fitting to every line of the stellar continuum subtracted long-slit spectroscopy image shown in Figure \ref{fig:2dspec_orig}. 
The pixel sampling of $0.2''$ corresponds to a physical scale of 1.3 kpc at the object. 
The radial distributions of flux (left), velocity shifts (middle) and FWHM (right) for each emission line are shown in Figure \ref{fig:2dspec_dist_orig}. 
The \oii\ line shows a more extended structure than stellar continuum, while \oiii\ and \ha\ emission lines show relatively compact flux distribution. 
The velocity profiles show an extent to about 5 kpc with velocity shift larger than 500 \kms\ and FWHM larger than 1000 \kms, especially in the North-East side, indicating the outflow is extended up to 5 kpc scale. However, there also exists the possibility that the observed \oiii\ line with high-velocity at 5 kpc comes from the smeared-out flux of an unresoved core component, e.g., compact star-forming region or AGN NLR, by the PSF, which could lead to an overestimation of the size of outflow. 

\begin{figure*}[!ht]
	\begin{center}
	\includegraphics[trim=0 10pt 0 0, width=\linewidth]{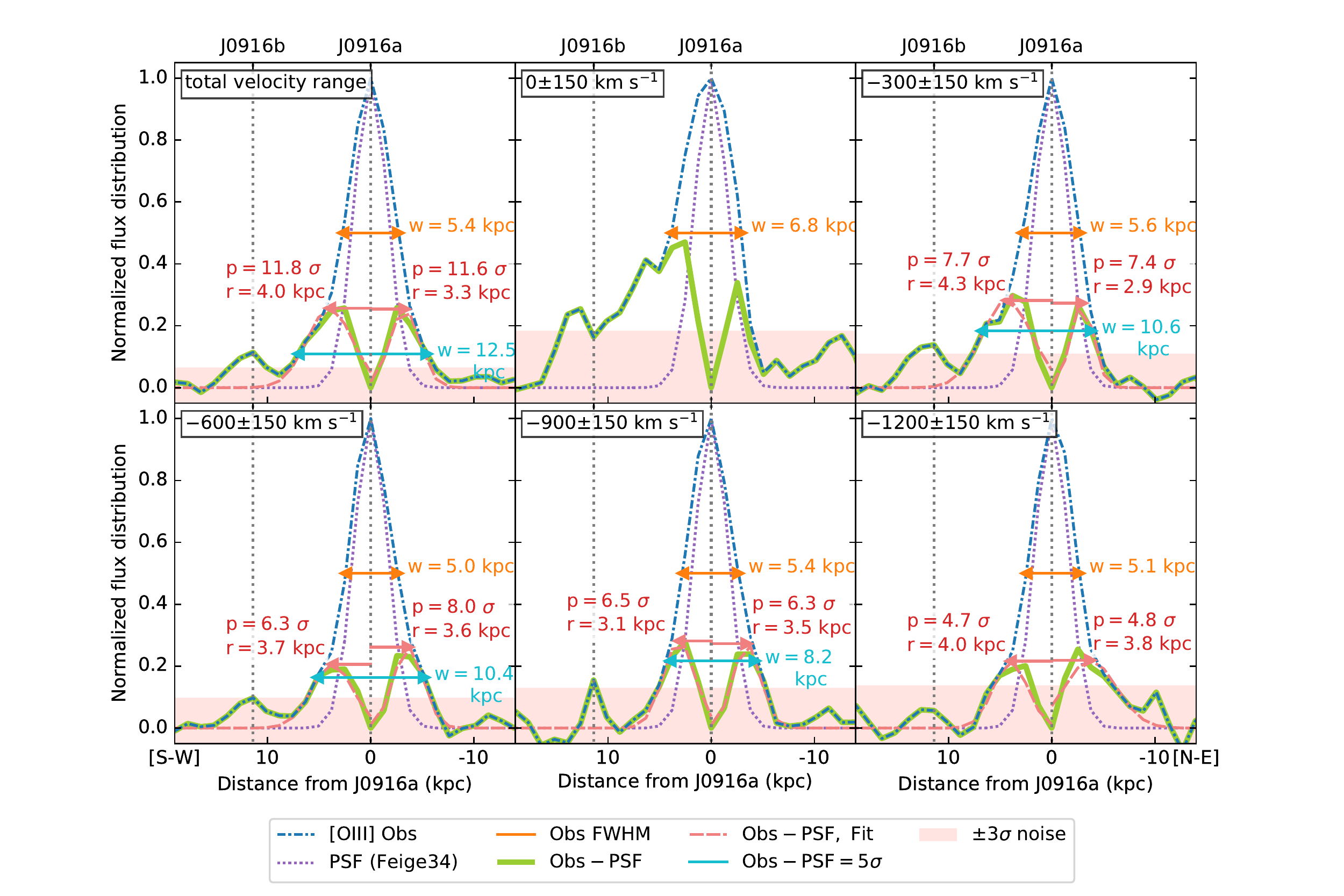}
	\end{center}
	\caption{
	The flux radial distribution of \oiii\ emission line per velocity window with a width of 300 \kms. The observed \oiii\ profile and the PSF from the observation of standard star Feige34 are shown with blue dash-dotted and violet dotted lines, respectively. The FWHM of the observed \oiii\ profile is shown in orange. 
	The unresolved core component is assumed to follow the profile of PSF and contribute to 100\% of the observed \oiii\ flux in the central position. The residuals of observed \oiii\ flux minus the PSF-convolved core component, which traces the extended \oiii\ flux, is shown with green solid curves. 
	We fit the residuals using one Gaussian profile on each of the South-West and North-East side. 
	The S/N ratios and radii of the peak of each Gaussian profile are marked in red. 
	The blue lines denote the outflow radius where the extended components can be detected at $5\sigma$ level, while the pink shadows indicate the $\pm3\sigma$ size. 
	The noise ($\sigma$) is estimated to be the scatter of the off-source region in the spectroscopy 2D image.
	}
	\label{fig:psf_ext}
\end{figure*}

In order to evaluate a possible effect of an unresolved component on the estimation of the size of the outflow region, 
firstly we plot the flux radial distribution per velocity window with a width of 300 \kms\ (Figure \ref{fig:psf_ext}) to check if high-velocity \oiii\ line can be detected in the outskirt region. The unresolved core component is assumed to follow the profile of PSF and contribute to 100\% of the observed \oiii\ flux in the central position. The residuals of observed \oiii\ flux minus the PSF-convolved core component (green solid curves in Figure \ref{fig:psf_ext}) can be used to trace the extended component. 
We fit the residuals using one Gaussian profile on each of the South-West and North-East side, and find that the extended component with a velocity shift of 600--900 \kms\ can be detected to about 3.5 kpc with S/N\,$\sim7$. The noise ($\sigma$) is estimated to be the scatter of the off-source region in the 2D spectroscopy image. If we take $5\sigma$ as the detection limit, then the maximum radius of the extended component is 4--5 kpc with a velocity shift of 600--900 \kms. 

\begin{figure*}[!ht]
	\begin{center}
	\includegraphics[trim=0 12pt 0 0, width=\linewidth]{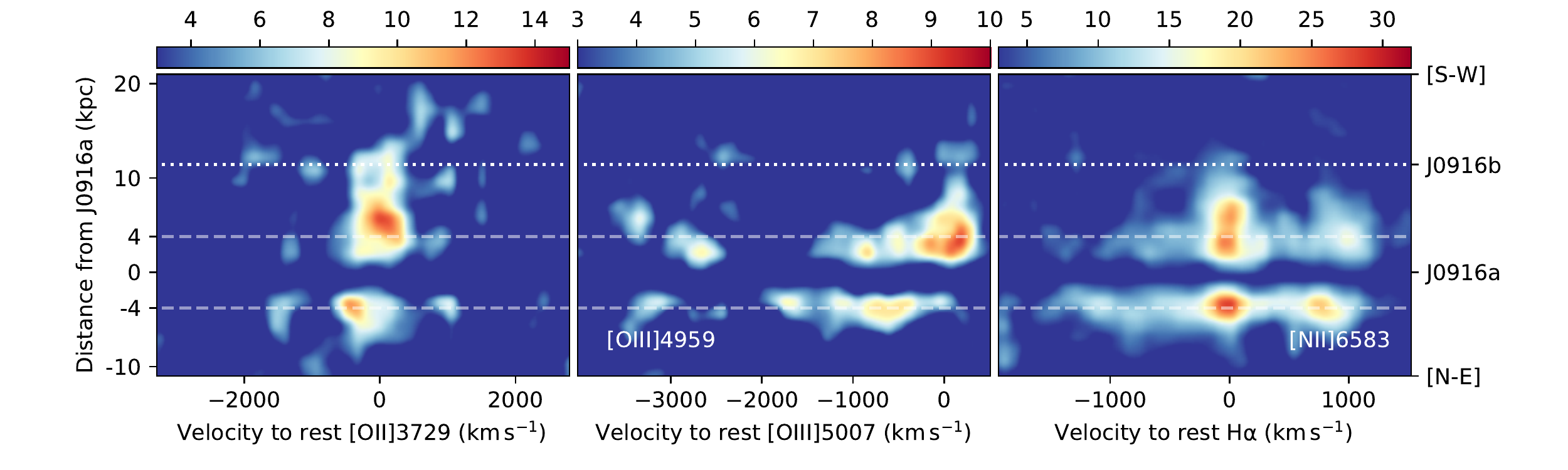}
	\end{center}
	\caption{
	Stellar continuum subtracted long-slit spectroscopy 2D image of extended \oii\ (left), \oiii\ (middle), and \ha-\nii\ (right) emission lines after removal of the unresolved core component. 
	The color bars represent the S/N ratios and only regions with S/N\,$>3\sigma$ are shown in each panel. The noise ($\sigma$) is estimated to be the scatter of the off-source region in the spectroscopy 2D image. 
	The 4 kpc outflow radius is shown with white dashed lines. 
	The velocity of \oii, \oiii\ and \nii\ emission lines on North-East side is higher than that on the South-West side, possibly indicating an inclined structure.  
    Other legends are the same as those in Figure \ref{fig:2dspec_orig}. }
    \label{fig:2dspec_deb}
\end{figure*}

\begin{figure*}[!ht]
	\begin{center}
	\hspace{-40pt}
	\includegraphics[trim=0 16pt 0 0, width=0.333\linewidth]{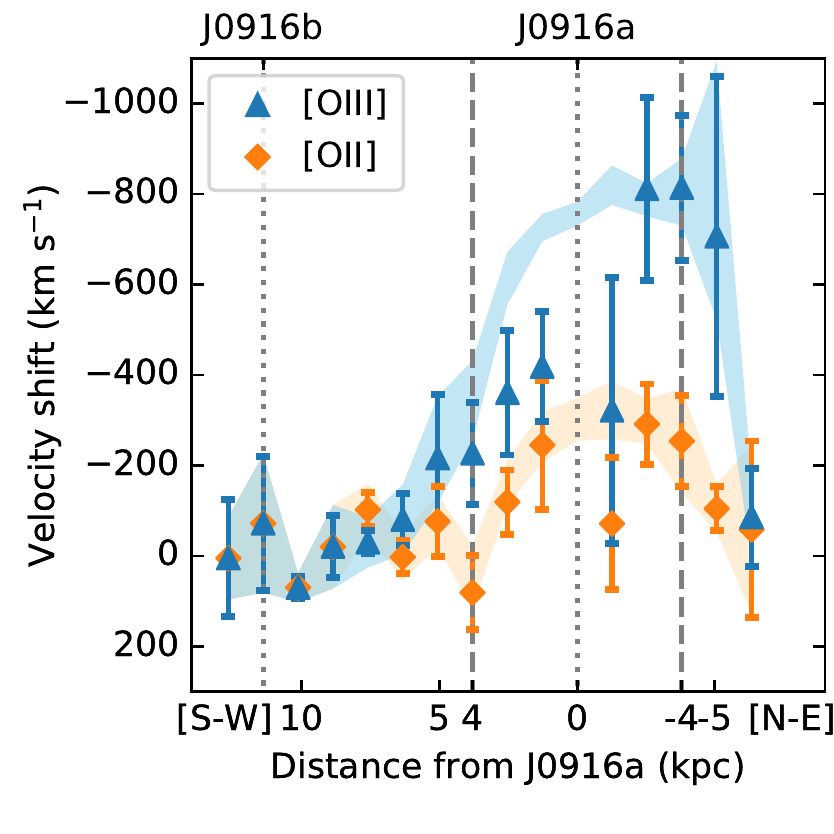}
	\includegraphics[trim=0 16pt 0 0, width=0.333\linewidth]{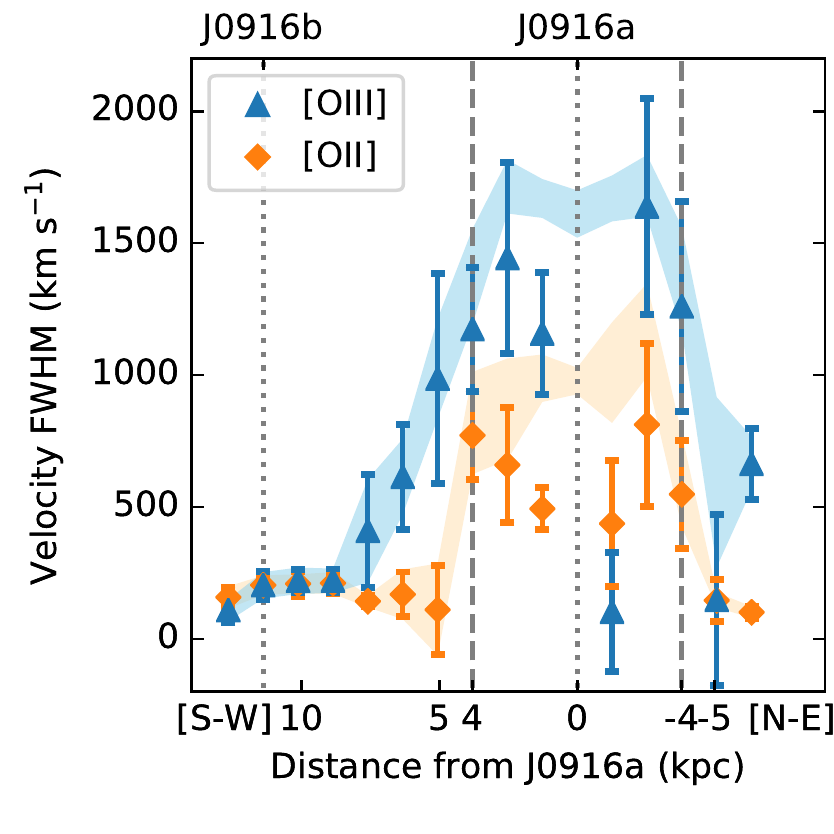}
	\end{center}
	\caption{
	The velocity shifts (left) and FWHM (right) along the slit direction of extended \oiii\ (blue triangles) and \oii\ (orange diamonds) emission lines and after removal of the PSF-convolved core component. 
	For the sake of comparison, the velocity shifts and FWHM before removing core component are also plotted using blue (\oiii) and orange (\oii) shadows. 
    }
	\label{fig:2dspec_dist_deb}
\end{figure*}

\begin{figure*}[!ht]
	\begin{center}
	\includegraphics[trim=56pt 30pt 56pt 40pt, width=0.32\linewidth]{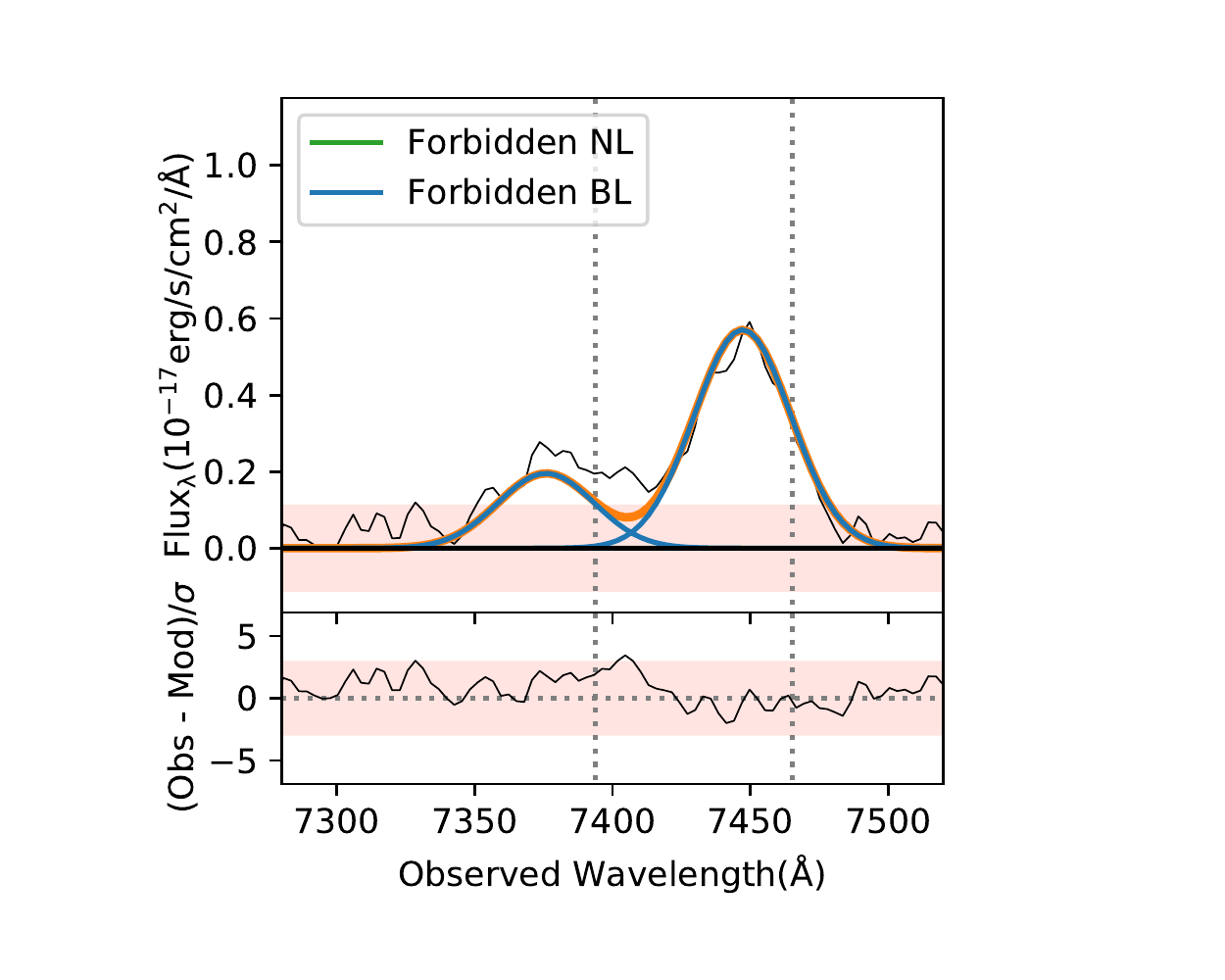}
	\includegraphics[trim=56pt 30pt 56pt 40pt, width=0.32\linewidth]{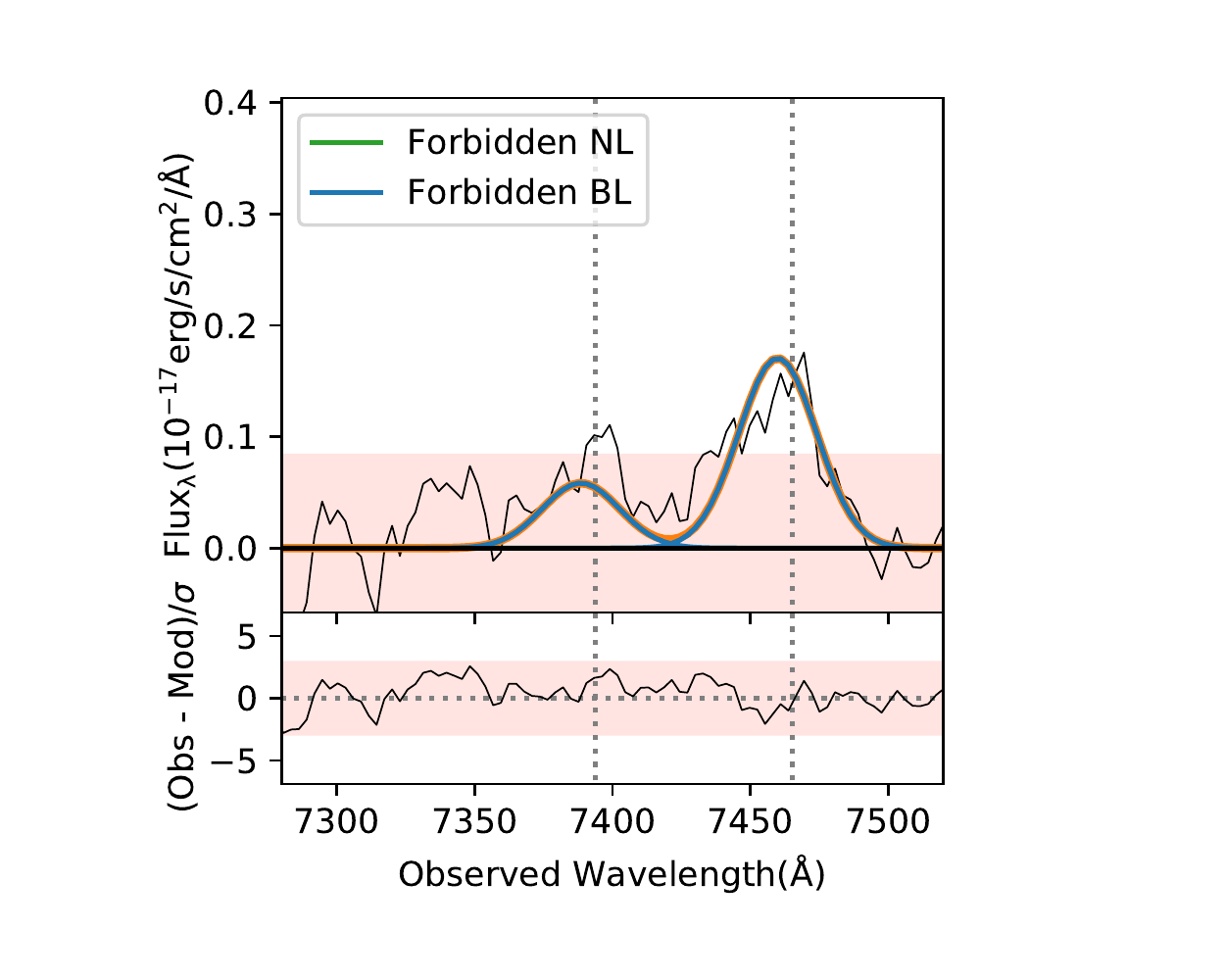}
	\includegraphics[trim=56pt 30pt 56pt 40pt, width=0.32\linewidth]{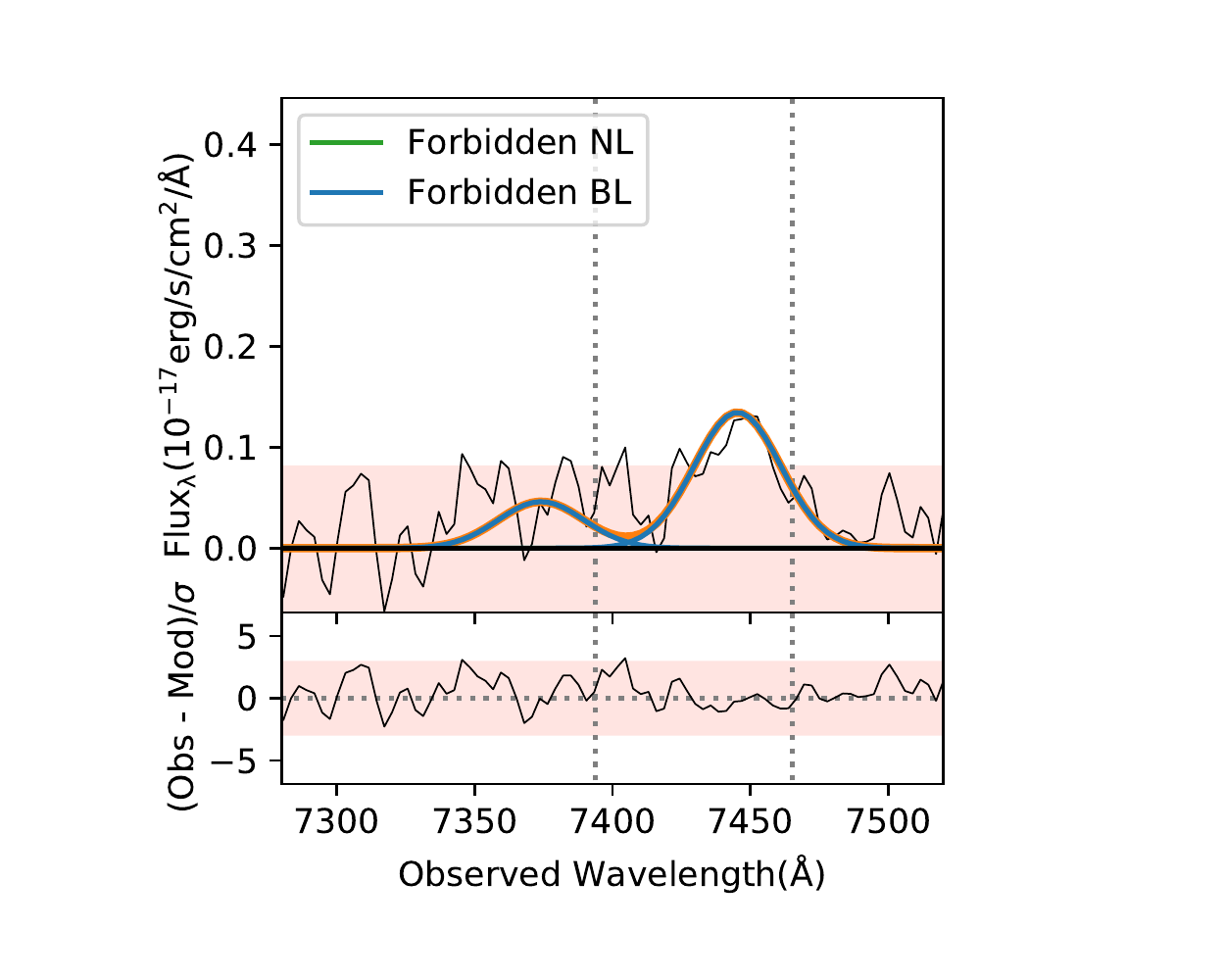}\\
	\end{center}
	\caption{
	\textbf{Left}: The \oiii\ emission lines extracted from the central pixel before the removal of PSF-convolved core component. 
	\textbf{Middle}: The \oiii\ emission lines extracted from the pixel 4 kpc from the center on South-West side, after the removal of PSF-convolved core component. 
	\textbf{Right}: The \oiii\ emission lines extracted from the pixel 4 kpc from the center on North-East side, after the removal of PSF-convolved core component. 
	The pink shadows mark the $3\sigma$ level. The noise ($\sigma$) is estimated to be the scatter of the line-free regions in each spectrum. 
	}
	\label{fig:2dspec_4kpc}
\end{figure*}

Assuming the original observed spectrum of the central pixel as the representative spectrum of the unresolved core component, we further subtract the PSF-convolved core component from the spectroscopy 2D image. The flux of the central pixel is entirely reduced. The spectroscopy 2D image of extended \oii, \oiii, and \ha-\nii\ emission lines after removal of the PSF-convolved core component are shown in Figure \ref{fig:2dspec_deb}. We re-perform spectral fitting per pixel along the long-slit direction, and the measured velocity shifts and FWHM of \oiii\ and \oii\ lines are shown in Figure \ref{fig:2dspec_dist_deb}. The results indicate that, even after the removal of the PSF-convolved core component, the \oiii\ line with high velocity shift ($\sim$\,800 \kms) and FWHM ($\sim$\,1200 \kms) can still be detected to 4 kpc scale on the North-East side. As for the South-West side, the velocity shift of \oiii\ decreases from $\sim$\,600 \kms\ to $\sim$\,300 \kms\ at 2--4 kpc. The difference between the velocity shift on North-East side and South-West side would imply that the outflow is within an inclined structure. Although the velocity shift of \oiii\ is lower on the South-West side, the FWHM of \oiii\ is still over 1000 \kms up to 4 kpc. Figure \ref{fig:2dspec_4kpc} shows the \oiii\ spectra of the unresolved core component (left), and the extended spectra after removal of the core component at 4 kpc on both sides (middle and right), respectively.  

The more extended flux distribution as well as the lower ionization potential of \oii\ suggest that the outflow shown by \oii\ emission line can be extended farther than \oiii\ line. 
However, Figure \ref{fig:2dspec_dist_deb} shows that the profiles of velocity shifts and FWHM of \oii\ line is more compact than those of \oiii. 
Furthermore, the velocity FWHM of extended \oii\ gas at 3--4 kpc decreases from 1000 \kms to 600 \kms after the removal of unresolved core component, indicating that the highest-velocity outflowing \oii\ gas is dominated by the core spectrum. 
Therefore, although the \oii\ gas is more spatially extended, the compact velocity profiles and core-dominated high-velocity outflowing gas imply that the extended \oii\ gas in the outer region of the galaxy has been less affected by the outflow than \oiii\ emitting gas.

\begin{figure}[!ht]
    \begin{center}
    \hspace{-11pt}
    \includegraphics[trim=0 30pt 0 0, width=0.5\textwidth]{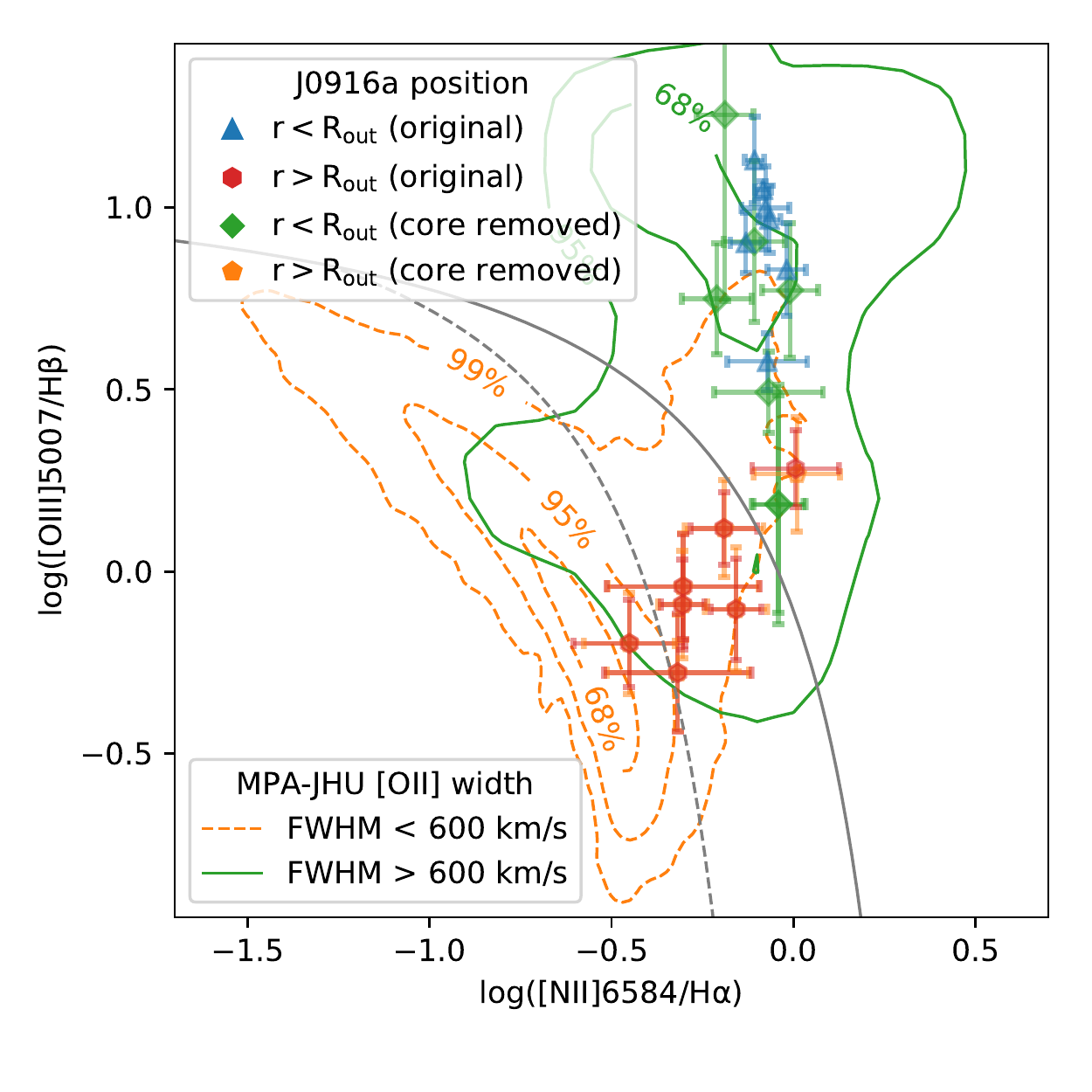}
    \end{center}
    \caption{
    BPT diagram of J0916a with the SDSS galaxies with low ($<600$ \kms, orange dashed contours) and high ($>600$ \kms, green solid contours) \oii\ velocity width. 
    The emission line ratios of J0916a calculated within the 4 kpc outflow region are shown in blue (original) and green (core component removed), while the red and orange markers show outside of the outflow region. 
    AGN, composite and star-forming galaxy regions are separated theoretically by solid (Kewley et al. 2001) and empirically by dashed (Kauffmann et al. 2003) black curves. 
    }
    \label{fig:bpt}
\end{figure}

The \nii\,/\,\ha\ and \oiii\,/\,\hb\ emission line ratios within the outflow region ($r<4$\, kpc) are higher than those of outside region. If we plot the line ratios of J0916a on the BPT diagram, then all the data points within outflow region are located on AGN (Seyfert 2) sequence (Figure \ref{fig:bpt}). 
It is widely accepted that the high-velocity outflows shown by emission lines with high-IP, e.g., \oiii\ and \neiii, are driven by AGN radiation or wind. 
On the other hand, the low-IP (13.62 eV) \oii\ line can also be significantly contributed from the star-forming regions, thus it is interesting to examine whether the fast \oii\ outflows are associated with the star formation activity. 
We use the MPA-JHU SDSS galaxy catalog to find the possible ionization source of fast \oii\ outflows observed among them. 
The distributions of galaxies with low and high FWHM of \oii\ emission line on the BPT diagram are shown in Figure \ref{fig:bpt}. 
Galaxies with FWHM larger than 600 \kms\ tend to show higher \nii\,/\,\ha\ and \oiii\,/\,\hb\ line ratios, and mostly lie on the AGN sequence, suggesting that the fast \oii\ outflowing gas is ionized by AGN radiation. 

\subsection{Mass outflow and energy ejection rates}\label{sec:discussion_ek_out}

In the recent works, the mass of the ionized outflowing gas has been estimated by either Balmer lines (\ha, e.g., \cite{Soto2012}, \cite{Arribas2014}; \hb, e.g., \cite{Harrison2014}, \cite{Carniani2015}) or \oiiiblong\ line (e.g., \cite{Cano2012, Kakkad2016}). Since the \hb\ emission of J0916a is weak, we estimate the mass of the ionized outflowing gas of J0916a using \ha\ and \oiii\ lines. 
Following \citet{Carniani2015}, the mass of the ionized gas can be expressed as:
\begin{eqnarray}\label{equ:mass_dev}
    M_{\rm ion} &\simeq& \int_V \left( m_{\rm p}n_{\rm H {\scriptscriptstyle II} } + 4 m_{\rm p}n_{\rm He {\scriptscriptstyle III} } \right) f_{\rm H {\scriptscriptstyle II} } \;\mathrm{d} V \nonumber \\
    &\simeq& m_{\rm p} 
    \left[ 1+4 \frac{n_{\rm He {\scriptscriptstyle III} }}{n_{\rm H {\scriptscriptstyle II} }} \right] 
    \left[\frac{n_{\rm H {\scriptscriptstyle II} }}{n_{\rm e}}\right] 
    \langle n_{\rm e} \rangle f_{\rm H {\scriptscriptstyle II} } V, 
\end{eqnarray} 
where $m_{\rm p}$ is the proton mass; $n_{\rm He {\scriptscriptstyle III} }$, $n_{\rm H {\scriptscriptstyle II} }$, and $n_{\rm e}$ are the number density of helium ions, hydrogen ions, and electrons, respectively; 
$V$ is the volume of the considered region and $f_{\rm H {\scriptscriptstyle II} }$ the volume filling factor of the ionized hydrogen gas. 
With Equation \ref{equ:mass_dev} and the luminosity of \ha\ emission line:
\begin{eqnarray}\label{equ:lum_dev_ha}
    L_{\rm H\alpha} &=& \int_V j_{\rm H\alpha}(n_{\rm e}, T_{\rm e})\, n_{\rm e} n_{\rm H {\scriptscriptstyle II} } f_{\rm H {\scriptscriptstyle II} } \;\mathrm{d} V \nonumber \\
    &\simeq& j_{\rm H\alpha} 
    \left[\frac{n_{\rm H {\scriptscriptstyle II} }}{n_{\rm e}}\right] 
    \langle n_{\rm e}^2 \rangle f_{\rm H {\scriptscriptstyle II} } V, 
\end{eqnarray} 
the mass of the ionized gas can be estimated as: 
\begin{equation}\label{equ:Mout_ha}
    \frac{ M_{\rm ion} }{10^8\ \rm M_{\odot} } = 3.330\times \frac{L_{\rm H\alpha}}{10^{43}\ \rm {erg\ s^{-1}} } \left( \frac{n_{\rm e}}{100\ \rm{cm^{-3}} } \right)^{-1}.
\end{equation} 
In the derivation we assume 
(a) helium gas is fully ionized and $n_{\rm He {\scriptscriptstyle III} }=0.1\times n_{\rm H {\scriptscriptstyle II} }$, thus $n_{\rm e}=n_{\rm H {\scriptscriptstyle II} } + 2 n_{\rm He {\scriptscriptstyle III} }=1.2\times n_{\rm H {\scriptscriptstyle II} }$; 
(b) $\langle n_{\rm e}^2 \rangle=\langle n_{\rm e} \rangle ^2$, i.e., all of the ionized gas clouds have the same density \citep{Cano2012}. 
The emissivity $j_{\rm H\alpha}=3.536\times10^{-25}$ \lumcgs ~cm$^{-3}$  is calculated using \texttt{PyNeb} \citep{Luridiana2015} under the typical temperature ($T_{\rm e}=10^4$ K) and electron density ($n_{\rm e}=100$ cm$^{-3}$) of \hii\ regions and AGN NLRs.
The mass of the ionized gas can be similarly derived from \oiii\ emission line, whose luminosity can be given by: 
\begin{eqnarray}\label{equ:lum_dev_oiii}
    L_{\rm [O {\scriptscriptstyle III}]} &=& \int_V j_{\rm [O {\scriptscriptstyle III}]}(n_{\rm e}, T_{\rm e})\, n_{\rm e} n_{\rm O \scriptscriptstyle III} f_{\rm O {\scriptscriptstyle III} } \;\mathrm{d} V \nonumber \\
    &\simeq& j_{\rm [O {\scriptscriptstyle III}]} 
    \left[\frac{n_{\rm O \scriptscriptstyle III}}{n_{\rm Oion}}\right] 
    \left[\frac{n_{\rm Oion}}{n_{\rm H {\scriptscriptstyle II} }}\right] 
    \left[\frac{n_{\rm H {\scriptscriptstyle II} }}{n_{\rm e}}\right] 
    \langle n_{\rm e}^2 \rangle f_{\rm O {\scriptscriptstyle III} } V, 
\end{eqnarray} 
where $n_{\rm O \scriptscriptstyle III}$ is the density of $\mathrm{O^{2+}}$ ions and $n_{\rm Oion}$ the total density of oxygen ions. 
In addition to the assumptions (a)-(b) adopted in the calculation with \ha, following \citet{Cano2012} and \citet{Carniani2015} we also assume 
(c) $n_{\rm Oion}=n_{\rm O \scriptscriptstyle III}$, i.e., most of the oxygen ions are in its doubly ionized form\footnote{If we consider the observed flux ratio of \oiii\ and \oii\, the actual $n_{\rm O \scriptscriptstyle III}/n_{\rm Oion}$ is approximately 0.7, which will result in a larger estimation of ionized gas mass. In order to compare the results with those in the literature (e.g., \cite{Harrison2012, Kakkad2016, Zakamska2016}), we keep $n_{\rm Oion}=n_{\rm O \scriptscriptstyle III}$ in the proceeding calculations.}; 
(d) $n_{\rm Oion}/n_{\rm H {\scriptscriptstyle II} }=n_{\rm O}/n_{\rm H}$, i.e., oxygen and hydrogen gas have the same ionization degree, and $n_{\rm O}/n_{\rm H}=[n_{\rm O}/n_{\rm H}]_{\odot}=730\pm100\ \mathrm{parts\ per\ million}$, where $[n_{\rm O}/n_{\rm H}]_{\odot}$ is the solar oxygen abundance \citep{Centeno2008}; 
(e) $f_{\rm O {\scriptscriptstyle III} } V=f_{\rm H {\scriptscriptstyle II} } V$, i.e., the regions from which oxygen and hydrogen lines are emitted have the same size. 
The emissivity $j_{\rm [O {\scriptscriptstyle III}]}=3.497\times10^{-21}$ \lumcgs ~cm$^{-3}$ is calculated with \texttt{PyNeb} under the same condition as that assumed for \ha\ ($n_{\rm e}=100$ cm$^{-3}$, $T_{\rm e}=10^4$ K). Combining Equation \ref{equ:mass_dev} and Equation \ref{equ:lum_dev_oiii}, we obtain the mass of the ionized gas as:
\begin{equation}\label{equ:Mout_oiii}
    \frac{ M_{\rm ion} }{10^8\ \rm M_{\odot} } = 4.613\times  \frac{L_{\rm [O {\scriptscriptstyle III}]} }{10^{44}\ \rm {erg\ s^{-1}} }  \left( \frac{n_{\rm e}}{100\ \rm{cm^{-3}} } \right)^{-1}.
\end{equation}

The mass of the ionized outflowing gas is proportional to the intrinsic luminosity of outflowing components of \ha\ and \oiii, and inversely proportional to the electron density. 
In order to obtain the intrinsic luminosity, the amount of dust extinction need to be determined. 
The color excess $E(B-V)$ can be estimated from the \ha\,/\,\hb\ emission line ratio through the formula:
\begin{equation}
    E(B-V)=-\frac{2.5}{k_{\rm H\alpha}-k_{\rm H\beta}}\log_{10} \left[ \frac{L_{\rm H\alpha,obs}/L_{\rm H\beta,obs}}{L_{\rm H\alpha,int}/L_{\rm H\beta,int}} \right],
\end{equation} 
where the attenuation curve, $k_{\lambda}=A_\lambda/E(B-V)$, which is calculated empirically for star-forming galaxies, resulting in $k_{\rm H\alpha}=3.33$ and $k_{\rm H\beta}=4.60$ \citep{Calzetti2000}. 
We assume the intrinsic value of the Balmer decrement, $L_{\rm H\alpha,int}/L_{\rm H\beta,int}$, of 2.86, which corresponds to the electron temperature of $10^4$ K and electron density of 100 cm$^{-3}$ assuming Case B recombination \citep{Osterbrock2006}. The ratio is commonly assumed for star-forming galaxies in the literature \citep{Groves2012, Dominguez2013}. 
Assuming the intrinsic value, we derive the color excess as $E(B-V)=1.01\pm0.30$ from the narrow components of \ha\ and \hb, which is close to the typical amount of dust attenuation in the local U/LIRGs \citep{Veilleux1995, Alonso2006, Garca2009} and DOGs \citep{Hwang2013}. 
The SED fitting yields higher color excess $E(B-V)_{\rm YSP}=1.49\pm0.10$ for the young stellar population. 
The structure of the dust extinction in ULIRGs was found to be patchy and clumpy \citep{Garca2009, Piqueras2013}, 
and $E(B-V)_{\rm YSP}=1.49$ possibly reflects the dust attenuation in dense molecular clouds where the young stars are formed. 
Since the relationship between the dust extinction of the stellar continuum and emission lines has not been concluded yet \citep{Puglisi2016, LoFaro2017}, we adopt $E(B-V)=1.01$ from the Balmer decrement to correct for the dust extinction of the outflowing gas, with the formula:  
\begin{equation}
    L_{\rm cor} = L_{\rm obs}\times 10^{0.4\times k_{\lambda}\times E(B-V)}.
\end{equation} 

Considering the projection effect in the line of sight, and 
for the comparison to the results in the literature, the total luminosity of \oiii\ including both of the unresolved core component and extended component is used to estimate the mass of the outflowing gas
\footnote{The integrated \oiii\ flux at 2.5--6.5 kpc contributes to approximately 35\% of the total \oiii\ flux of the galaxy. The estimated gas mass will decrease by 0.45 dex if only the gas in the spatially extended component is considered in the calculation.}. 
The luminosity of broad \ha\ is obtained from the spectral fitting decomposition with both of the broad \ha\ and broad \nii\ components (see discussion in Section \ref{sec:spectral_analyses}). 
The \siilong\ doublet of J0916a lies in the observed wavelength range where the contamination from night sky lines is severe, thus the electron density is assumed to be $n_{\rm e}=100$ cm$^{-3}$, which is a conventionally employed value for the estimation of ionized outflowing gas mass \citep{Liu2013, Kakkad2016, Toba2017}. 
With the dust extinction corrected luminosity and the assumption of $n_{\rm e}=100$ cm$^{-3}$, the mass of the ionized outflowing gas in J0916a is estimated to be $M_{\rm out, H\alpha}=4.7\times10^{8}$ \ms\ using the broad component of \ha, or $M_{\rm out, [O {\scriptscriptstyle III}]}=3.1\times10^{8}$ \ms\ using the \oiii\ emission line. 
The $M_{\rm out, H\alpha}/M_{\rm out, [O {\scriptscriptstyle III}]}\sim1.5$ is consistent with the average ratio ($\sim3$) of AGN outflows \citep{Fiore2017}. The difference between $M_{\rm out, H\alpha}$ and $M_{\rm out, [O {\scriptscriptstyle III}]}$ can be due to the assumptions employed in Equation \ref{equ:Mout_oiii}, e.g., $n_{\rm Oion}=n_{\rm O \scriptscriptstyle III}$ possibly results in the underestimation of $M_{\rm out, [O {\scriptscriptstyle III}]}$. 

In gas clouds with constant density, 
the mass outflow rate can be derived as $\dot{M}_{\rm out}=M_{\rm out}\times\dot{V}/V=M_{\rm out}v_{\rm out}\times A/V$, 
where $v_{\rm out}=w_{80}/1.3$ is the bulk outflow velocity \citep{Liu2013}, 
and $A/V$ is the surface-area-to-volume ratio. 
Assuming the ionized outflowing gas in a spherically symmetric sector \citep{Fiore2017}, we have $A/V=\Omega R_{\rm out}^2/(\Omega R_{\rm out}^3/3)=3/R_{\rm out}$, where $\Omega$ is the opening angle and the $R_{\rm out}$ is the radius of outflow region. 
We adopt $R_{\rm out}\sim4$ kpc, which is estimated from the core-component-removed spectroscopy 2D image at $5\sigma$ level. 
Then the mass outflow rate $\dot{M}_{\rm out}$ and energy ejection rate $\dot{E}_{\rm out}$ can be derived as:
\begin{equation}\label{equ:Eout_sphere}
    \dot{M}_{\rm out} = M_{\rm out}v_{\rm out}\times\frac{3}{R_{\rm out}} , \ \
    \dot{E}_{\rm out} = \frac{1}{2}\dot{M}_{\rm out} v_{\rm out}^2 ,
\end{equation} 
with the results of 
$560$ \sfrunit\ and $4.1\times10^{44}$ \lumcgs\ using broad \ha\ line, 
or
$370$ \sfrunit\ and $2.7\times10^{44}$ \lumcgs\ using \oiii\ line,
respectively. Those values reflect the instantaneous feedback effect from AGN and\,/\,or star formation activity in the radius $R_{\rm out}$ at the observed time. 

The estimation of $\dot{M}_{\rm out}$ and $\dot{E}_{\rm out}$ is based on a few assumptions and the results are dependent on the definition of outflow velocity and kinetic power, the size of outflow region, 
the correction for the PSF smearing effect, 
the correction for dust extinction, as well as the electron density. 
The derived $\dot{E}_{\rm out}$ are consistent with those estimated with another conventional definition of kinetic power 
$\dot{E}_{\rm out} = \dot{M}_{\rm out} \left( \Delta v^2+3 \sigma ^2 \right)/2$ \citep{Holt2006, Harrison2014}, 
where $\Delta v = (v_{05}+v_{95})/2$ ($v_{05}$ and $v_{95}$ are defined from Equation \ref{equ:w80}) and $\sigma=w_{80}/2.355$, 
within a factor of 0.6. 
\citet{Bae2017} proposed an outflow velocity estimate as $v_{\rm out}^{\star}\equiv 2\times\sqrt{v^2+\sigma^2}$, where $v$ and $\sigma$ are the flux-weighted averaged shift and dispersion of the emission lines. 
If we calculate $\dot{M}_{\rm out}$ and $\dot{E}_{\rm out}$ with the definition of $v_{\rm out}^{\star}$, the results increase by a factor of 1.5. 
$\dot{M}_{\rm out}$ and $\dot{E}_{\rm out}$ will decrease by 0.45 dex if only the gas in the spatially extended component is considered in the calculation. 
The uncertainty in the extinction correction using $E(B-V)=1.01\pm0.30$ can lead to the variations of $\dot{M}_{\rm out}$ and $\dot{E}_{\rm out}$ by 0.4 and 0.5 dex for \ha\ and \oiii\ lines, respectively.
The assumption of the electron density, 
which is thought as the main source of uncertainties in the calculation of ionized gas mass and hence the kinetic power of the outflow, was widely discussed in the literature \citep{RodriguezZ2013, Fiore2017, Harrison2018}. 
The typical value of $n_{\rm e}$ for the AGN NLR is 50--1500 cm$^{-3}$ \citep{Peterson1997}. 
100--500 cm$^{-3}$ was assumed in several papers \citep{Cano2012, Harrison2014, Carniani2015, Kakkad2016, Fiore2017,Toba2017} and roughly consistent with the result of \sii\ ratio of a luminous obscured quasar \citep{Perna2015}. 
\citet{Rupke2013} and \citet{Liu2013} employed a lower density of 10 cm$^{-3}$ for ionized outflowing gas clouds in a thin shell, which was found in the superbubble in NGC 3079 \citep{Cecil2001}. 
On the contrary, \citet{Holt2011}, \citet{Rose2018}, and \citet{Kawaguchi2018} reported that the density of outflowing gas can reach as high as $10^3$ to $10^5$ cm$^{-3}$. 
We have no direct constraint on $n_{\rm e}$ for J0916a. However, the critical density of \oii\ ($\sim10^{4}$ cm$^{-3}$, \cite{Osterbrock2006}) can be considered as a proxy for the upper limit of $n_{\rm e}$. 
The large uncertainty of $n_{\rm e}$ can result in the uncertainty of the estimated gas mass and kinetic power of the outflow by one to two orders of magnitude. 

A method assuming an energy-conserving bubble in a uniform medium was also widely used to estimate the maximum of energy ejection rates \citep{Heckman1990, Veilleux2005, Nesvadba2006, Harrison2012, Harrison2014}. 
In the scenario \citep{Castor1975, Weaver1977}, the fast inside wind induced by the central engine release energy into a hot ($T \ge 10^6$ K), low-density, and coronal shocked gas bubble, where the radiative cooling loss is insignificant, by a constant rate, $L_{\rm wind}$. 
The adiabatically expanding shocked bubble sweeps up a warm ($T \sim 10^4$ K), dense, and thin shell into the ISM in the host galaxy. 
The mechanical luminosity of the inside AGN or stellar wind, $L_{\rm wind}$, can be estimated with the formula:
\begin{equation}\label{equ:Eout_bubble}
    \frac{ L_{\rm wind} }{10^{46}\ \rm erg\ s^{-1}} = 3.0\times \Big( \frac{v_{\rm out}}{1000\ \rm km\ s^{-1}} \Big)^3 \Big( \frac{R_{\rm out}}{10\ \rm kpc} \Big)^2 \frac{n_0}{1\ \rm cm^{-3}},
\end{equation} 
where $n_0$ is the gas density of the un-disturbed ambient ISM, and we adopt $n_0=0.5$ cm$^{-3}$ following \citet{Nesvadba2006} and \citet{Harrison2014}; $R_{\rm out}$ and $v_{\rm out}$ are the same as defined in Equation \ref{equ:Eout_sphere}. According to the calculation of \citet{Weaver1977}, 55\,\% of the total feedback energy is released into the swept-up gas region, and 40\,\%--70\,\% of the shell energy is converted to the kinetic energy of the swept-up gas as it collapse into a thin shell through radiative cooling. 
Therefore the kinetic power of the swept-up ISM corresponds to 22\,\%--38\,\% of the feedback energy, i.e., $\dot{E}_{\rm out} \simeq 0.30 \times L_{\rm wind}$. 
For the sake of clarity, the kinetic energy ejection rates derived from the uniform spherical sector model and the energy-conserving bubble model, are named $\dot{E}_{\rm k,out}^{\rm sph}$ and $\dot{E}_{\rm k,out}^{\rm bub}$, respectively. 
The $\dot{E}_{\rm k,out}^{\rm bub}$ from \oiii\ lines is $2.6\times10^{45}$ \lumcgs, which should be considered as the upper limit of the kinetic energy ejection rate because in the scenario it is assumed that all the ambient ISM within $R_{\rm out}$ is entrained in the outflow and no filling factor is employed to reflect the clumpiness of the ambient gas.  
If we adopt filling factor of 0.2, which is a typical value for warm neutral medium (\hbox{H\sc i}) and warm ionized medium (\hbox{H\sc ii}) of the Galaxy \citep{Brinks1990}, the estimation of $\dot{E}_{\rm k,out}^{\rm bub}$ will decrease by approximately one order of magnitude. 

The derived mass outflow rate (400--500 \sfrunit, from the uniform spherical sector model) is comparable to the star formation rate (990 \sfrunit), leading to a mass loading factor of $\eta=\dot{M}_{\rm out}/\textrm{SFR}\sim0.5$, which informs the efficiency of the outflow to remove gas from star-forming region. 
The co-existence of the strong outflow and intense star formation indicates that the feedback of the outflow has not severely affect the star-forming region of the galaxy, although the outflow would be extend to the radius of 4 kpc. 
The probable explanation can be that the axis of the outflow is perpendicular to the disc of the host galaxy. 
It is also possible that the galaxy stays in the intermediate stage in which the feedback just becomes effective and begins to sweep out the ISM in the entire galaxy, i.e., on-set of a quenching of star formation.

In order to discuss the dominant power source of the outflow further, we can compare the energy input power from the AGN and star formation to the energy ejection rate \citep{Cicone2014, Kakkad2016, Harrison2018}. The coupling efficiency $\epsilon_{\rm f}$ is defined as the fraction of energy from AGN or star formation that couples to the surrounding gas. In the simulations of galaxy evolution with AGN feedback, $\epsilon_{\rm f,AGN}^{\rm m}$ is assumed to be from 0.5\,\% (assuming hot gas fraction of 10\,\%, \cite{Hopkins2010}) to 20\,\% (employed for AGN in the low-accretion mode, \cite{Weinberger2017}), with a typical value of 5\,\%, which is estimated in \citet{DiMatteo2005} to reproduce the normalization of the local $M_{\rm BH}$--$\sigma$ relation. As for feedback from star formation activity, the power provided by a starburst can be estimated with the formula, $P_{\rm SF}$ (\lumcgs) $=7\times10^{41}$ SFR (\sfrunit), by assuming that gas has a solar-metallicity and the mass-loss rate and mechanical luminosity are constant beyond 40 Myr \citep{Veilleux2005}. Combining with the empirical relationship of SFR (\sfrunit) $=4.5\times10^{-44}\ L_{\rm IR}^{\rm dust}$ (\lumcgs) \citep{Kennicutt1998}, we obtain $\epsilon_{\rm f,SF}^{\rm m}=P_{\rm SF}/L_{\rm IR}^{\rm dust}=3.15\,\%$. 
From the SED fitting, we obtain the AGN bolometric luminosity, $L_{\rm AGN}=1.78\times10^{45}$ \lumcgs, as well as the star formation contributed IR luminosity, $L_{\rm SF}=L_{\rm IR}^{\rm dust}=2.20\times10^{46}$ \lumcgs. Therefore the observed kinetic coupling efficiency is 
$\dot{E}_{\rm k,out}^{\rm sph}/L_{\rm AGN}\sim20\,\%$ and $\dot{E}_{\rm k,out}^{\rm sph}/L_{\rm SF}\sim2\,\%$, respectively. Hence, both AGN and star formation are sufficient to drive the strong outflow, but larger coupling efficiency would be required for the AGN feedback case. 

\begin{figure*}[!ht]
    \begin{center}
    \hspace{-20pt}
    \includegraphics[trim=0 8pt 0 0, width=\textwidth]{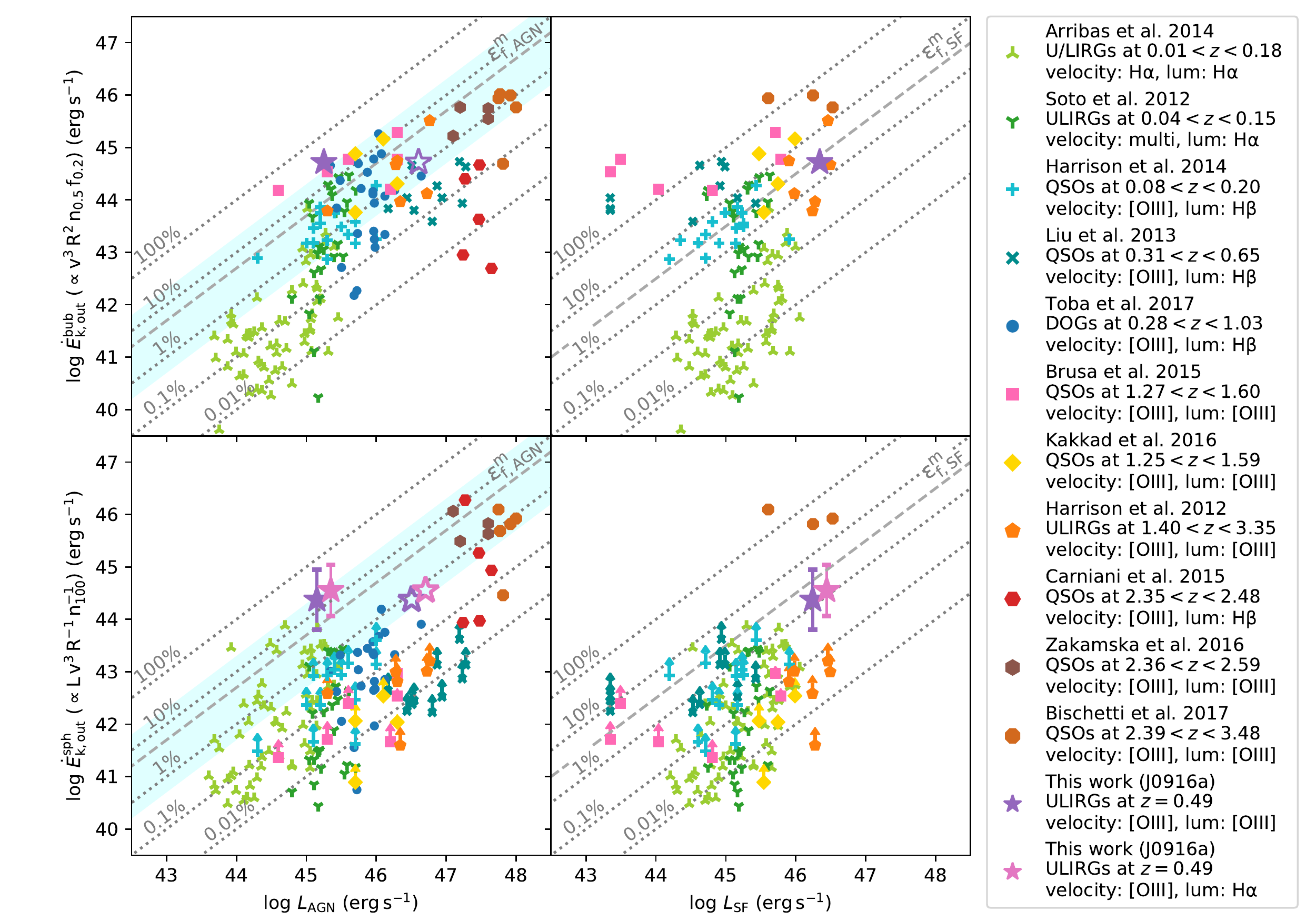}
    \end{center}
    \caption{
    \textbf{Bottom-left}: Energy ejection rates derived with the uniform spherical sector model, $\dot{E}_{\rm k,out}^{\rm sph}$, versus the AGN bolometric luminosities, $L_{\rm AGN}$. 
    \textbf{Bottom-right}: $\dot{E}_{\rm k,out}^{\rm sph}$ versus the star formation contributed IR luminosities $L_{\rm SF}$. 
    \textbf{Upper-left}: Energy ejection rates derived with the energy-conserving bubble model, $\dot{E}_{\rm k,out}^{\rm bub}$, versus $L_{\rm AGN}$.
    \textbf{Upper-right}: $\dot{E}_{\rm k,out}^{\rm bub}$ versus $L_{\rm SF}$. 
    The results of J0916a estimated using \oiii\ and \ha\ are shown with violet and purple filled stars, respectively. 
    The error bars indicate the scatters of $\dot{E}_{\rm k,out}^{\rm sph}$ ($\sim$ 0.5 dex) owing to the uncertainties of line fitting and extinction correction. 
    The open stars denote the $L_{\rm AGN}$ of J0916a estimated with the bolometric correction \citep{Kauffmann2009} to the extinction corrected \oiii\ luminosity. 
    The results of J0916a from \oiii\ and broad \ha\ lines are separated by an offset of 0.2 dex in the direction of x-axis. 
    For reference, we also plot the results of the local U/LIRGs and QSOs (\cite{Arribas2014, Soto2012, Harrison2014}), 
    the DOGs and QSOs at intermediated redshifts \citep{Liu2013, Toba2017, Brusa2015, Kakkad2016} 
    as well as the ULIRGs and QSOs at high redshifts (\cite{Harrison2012, Carniani2015, Zakamska2016, Bischetti2017}), 
    whose redshifts and emission lines used for estimating outflow velocities and luminosities are listed in the legend. 
    For the galaxies in all of the reference samples, the masses of ionized outflowing gas are estimated with Equation \ref{equ:Mout_ha} for Balmer lines or Equation \ref{equ:Mout_oiii} for \oiii\ line, then the energy ejection rates, $\dot{E}_{\rm k,out}^{\rm sph}$ and $\dot{E}_{\rm k,out}^{\rm bub}$, are re-calculated using Equation \ref{equ:Eout_sphere} and Equation \ref{equ:Eout_bubble}, respectively. 
    In the uniform spherical sector model, it is assumed that all of the galaxies have the same electron density ($n_{\rm e}=100$ cm$^{-3}$) and electron temperature ($T_{\rm e}=10^4$ K). 
    In the energy-conserving bubble model, we adopt the kinetic fraction of feedback energy, i.e., $\dot{E}_{\rm out} \simeq 0.30 \times L_{\rm wind}$, and employ a filling factor of 0.2 to reflect the clumpiness of the ambient gas. 
    In the case where the $w_{80}$ are not available, we use the $1.08\times\mathrm{FWHM}$ of the outflowing components instead (for a single Gaussian profile $w_{80}=1.08\times\mathrm{FWHM}$). 
    The objects in the samples of Harrison et al. (2012, 2014), \citet{Liu2013}, \citet{Brusa2015}, and \citet{Kakkad2016} only have lower limit of $\dot{E}_{\rm k,out}^{\rm sph}$ since the luminosities of outflowing gas were not corrected for dust extinction. 
    The dotted lines denote the energy ejection rate corresponds to 0.01\,\%--100\,\% of the AGN or star formation luminosity. 
    The dashed lines mark the feedback coupling efficiencies predicted by the models ($\epsilon_{\rm f,AGN}^{\rm m}$ = 5\,\%, \cite{DiMatteo2005}; $\epsilon_{\rm f,SF}^{\rm m}$ = 3.15\,\%, \cite{Veilleux2005}). The light cyan shadow shows the range of $\epsilon_{\rm f,AGN}^{\rm m}$ from 0.5\,\% \citep{Hopkins2010} to 20\,\% \citep{Weinberger2017}. 
    }
    \label{fig:comp_ek_out_quad}
\end{figure*}

In order to compare the energy ejection rate of J0916a to that reported in the literature, we collect the outflow samples 
from the local U/LIRGs and QSOs (\cite{Soto2012, Arribas2014, Harrison2014}), 
the DOGs and QSOs at intermediated redshifts \citep{Liu2013, Brusa2015, Kakkad2016, Toba2017}, 
to the ULIRGs and QSOs at high redshifts (\cite{Harrison2012, Carniani2015, Zakamska2016, Bischetti2017}). 
In majority of the samples, the $L_{\rm AGN}$ is estimated using multi-band SED fitting, except for 
\citet{Arribas2014}\footnote{The AGN contributions to the total luminosity of the galaxies in the sample of \citet{Arribas2014} were not presented, we apply an average AGN fraction of 20\,\% as mentioned in Section 4.2 of \citet{Arribas2014}.}, 
\citet{Liu2013}\footnote{Liu et al. (2013) estimated $L_{\rm AGN}$ using the rest 12 $\mu$m luminosity, $\nu L_{\rm 12 \mu m}$, and the bolometric correction of 9 from Richards et al. (2009).}, 
and \citet{Carniani2015}\footnote{\citet{Carniani2015} estimated $L_{\rm AGN}$ using the empirical relation $L_{\rm AGN}\simeq6\times \lambda L_{5100}$.}. 
For the galaxies in all of the reference samples, the masses of ionized outflowing gas are estimated using Equation \ref{equ:Mout_ha} for Balmer lines or Equation \ref{equ:Mout_oiii} for \oiii\ line, and then the energy ejection rates, $\dot{E}_{\rm k,out}^{\rm sph}$ and $\dot{E}_{\rm k,out}^{\rm bub}$, are re-calculated using Equation \ref{equ:Eout_sphere} and Equation \ref{equ:Eout_bubble}, respectively. 
For illustrative purposes, we assume that in the uniform spherical sector model, all of the galaxies have the same electron density ($n_{\rm e}=100$ cm$^{-3}$) and electron temperature ($T_{\rm e}=10^4$ K). 
In the energy-conserving bubble model, we adopt the kinetic fraction of feedback energy, i.e., $\dot{E}_{\rm out} \simeq 0.30 \times L_{\rm wind}$, and employ a filling factor of 0.2 to reflect the clumpiness of the ambient gas. 
The results are also shown in Figure \ref{fig:comp_ek_out_quad}. 

J0916a shows one of the strongest outflow among the galaxies at $z<1.6$. 
However, 
if we consider the bolometric luminosity of AGN ($L_{\rm AGN}$) from the SED fitting, 
the $L_{\rm AGN}$ of J0916a corresponds to only 1\,\%--10\,\% of $L_{\rm AGN}$ of the galaxies with similar energy ejection rate, which would suggest the largest coupling ratio ($\dot{E}_{\rm out}/L_{\rm AGN}$) among the plotted objects in Figure \ref{fig:comp_ek_out_quad}. 
In order to investigate the status of the AGN activity, we also estimate the Eddington ratio $\lambda_{\rm Edd}\equiv L_{\rm AGN}/L_{\rm Edd}$ of J0916a. 
The Eddington luminosity is defined with (e.g., \cite{Rybicki1979}): 
\begin{equation}
    \frac{L_{\rm Edd}}{10^{46}\ \rm erg\ s^{-1}} \equiv 1.26 \times \frac{M_{\rm BH}}{10^8\ \rm{M_{\odot}} }.
\end{equation} 
As for J0916a, the black hole mass $M_{\rm BH}$ can be estimated from the total stellar mass $M_{\star}$ with an empirical relationship 
between $M_{\rm BH}$ and $M_{\star}$ in the local universe \citep{Reines2015}:
\begin{equation}
    \log_{10}(\frac{M_{\rm BH}}{10^8\ \rm{M_{\odot}} }) = 1.05 \times \log_{10}(\frac{M_{\star}}{10^{11}\ \rm{M_{\odot}} })-0.55.
\end{equation} 
From the SED fitting we obtain the estimation of $M_{\star}=9.46\pm1.69\times10^{10}$ \ms. 
Hence the black hole mass and Eddington luminosity are derived as $M_{\rm BH}=2.66\times10^7$ \ms\ and $L_{\rm Edd}=3.35\times10^{45}$ \lumcgs, respectively. 
The Eddington ratio $\lambda_{\rm Edd}=0.6$ of J0916a lies between those of ULIRGs\,/\,DOGs, which show a super-Eddington accretion 
(3.35, \cite{Kawakatu2007}\footnote{Note that \citet{Kawakatu2007} calculated the ratio of $L_{\rm IR}^{\rm tot}/L_{\rm Edd}$ instead of $L_{\rm AGN}/L_{\rm Edd}$. The nuclear starbursts ($<1$ kpc) can contribute a fraction ($<30\,\%$) of the infrared luminosity for their ULIRG sample. }
; 2--3, \cite{Assef2015}; $\sim$ 3, \cite{Toba2017}), and nearby QSOs, which show a sub-Eddington value (0.09, \cite{Kawakatu2007}). 

\begin{figure}[!ht]
	\begin{center}
    \hspace{-11pt}
	\includegraphics[trim=0 24pt 0 0, width=0.5\textwidth]{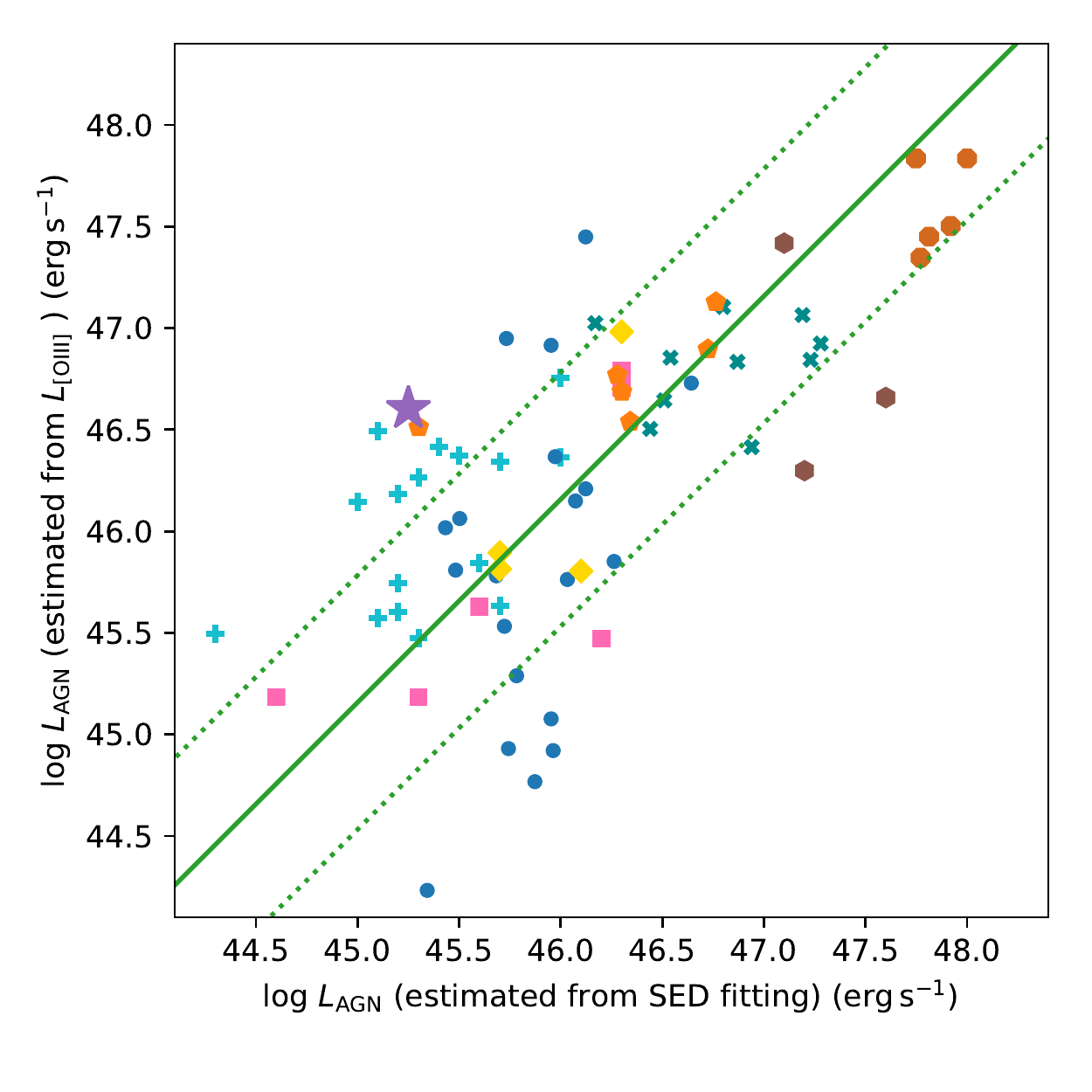}
	\end{center}
	\caption{
	The comparison between SED-based and \oiii-based AGN bolometric luminosity ($L_{\rm AGN}$). The ULIRG, J0916a, is shown with violet stars. The green solid and dotted lines denote the linear fitting result and $1\sigma$ range for the sources form the literature. Other legends are the same as those in Figure \ref{fig:comp_ek_out_quad}. 
	}
	\label{fig:comp_L_AGN}
\end{figure}

The extremely strong outflow associated with less luminous AGN in J0916a is puzzling. 
It is possible that the AGN bolometric luminosity ($L_{\rm AGN}$) of J0916a is underestimated in the SED fitting. 
We compare the SED-based $L_{\rm AGN}$ and \oiii-based $L_{\rm AGN}$ for this ULIRG and the reference sources in Figure \ref{fig:comp_ek_out_quad}, and the comparisons are shown in Figure \ref{fig:comp_L_AGN}. The bolometric correction factor of \oiii\ is 600 for extinction-corrected sources and 3500 for extinction-uncorrected sources \citep{Kauffmann2009, Lamastra2009}. 
The luminosity ratio, $\log{(L_{\rm AGN, [O {\scriptscriptstyle III}]}/L_{\rm AGN, SED})}$, is $0.2\pm0.6$ for the reference sources, and 1.3 for J0916a, indicating the $L_{\rm AGN}$ of J0916a estimated from the extinction-corrected \oiii\ luminosity is about 20 times higher than that estimated from the AGN-contributed IR luminosity. 
If we assume the \oiii-based $L_{\rm AGN}$, then J0916a (shown as open stars in Figure \ref{fig:comp_ek_out_quad}) becomes close to the most luminous high-$z$ QSOs with the most powerful outflows in the $\dot{E}_{\rm out}$--$L_{\rm AGN}$ diagram. 
The Eddington ratio $\lambda_{\rm Edd}\equiv L_{\rm AGN, [O {\scriptscriptstyle III}]}/L_{\rm Edd}\sim10$ is similar with that of the DOGs/ULIRGs with super-Eddington accretion \citep{Kawakatu2007, Assef2015, Toba2017}. 
The \oiii-based $L_{\rm AGN}$ requires that the entire IR luminosity is owing to AGN, which is hard to be reproduced by the current SED fitting templates, where the FIR peak is mainly contributed by star formation heated dust. The hard X-ray observation is necessary to determine the intrinsic power of the possible hidden AGN in J0916a. 

Another possible explanation for the large outflow coupling ratio is that J0916a is in the phase where the central AGN is less active than its peak epoch. 
The low IR luminosity of AGN, which originates from dusty torus in the vicinity (pc-scale) of SMBH, would imply the possibility that AGN lies in a fading status; while the observed extreme \oiii\ and \oii\ outflows would reflect a historical effect of the central engine during its preceding active phase, due to the time-lag between AGN activity in a nuclear region and outflow in an ionization cone (kpc-scale; \cite{Harrison2017}). 
The traveling timescale of the outflow in J0916a can be estimated with $R_{\rm out}/v_{\rm out}$, which results in $\sim3$ Myr. 
Theoretical simulations \citep{Novak2011, Gabor2014} and observational investigations \citep{Hickox2014} suggest that the accretion rates of SMBHs and AGN luminosities can vary by several orders of magnitude in timescales of $\sim1$ Myr or less. 
Additionally, recent works reported a population of AGNs called `dying AGNs', which show AGN signatures in large spatial scale (e.g., radio jets and\,/\,or bright \oiii\ line in the kpc-scale NLRs), but lack the features in small scale (e.g., weak or lack of X-ray and\,/\,or MIR emission), and imply the transient stage that the central engine was active in the past, but now seems quenched (\cite{Schawinski2010}; \cite[2017, 2019]{Ichikawa2016}; \cite{Schirmer2016}). Both AGN variability and AGN quenching can be one of possible mechanisms of the fading AGN scenario of J0916a. 

It is also possible that star formation activity at least partially contributes to the strong outflow. 
While several studies presented that AGNs are required to accelerate high-velocity winds (e.g., $v_{\rm max}=| v_{\rm shift} | + \textrm{FWHM}/2 \ge 500$ \kms, \cite{Rupke2005, Westmoquette2012, Arribas2014, Harrison2014}), recently high-velocity outflows traced using \hbox{Mg\sc ii 2796\AA\ 2803\AA} doublet absorption lines in galaxies with intense star formation activity but weak or no AGN activity have been reported \citep{Diamond2012, Bradshaw2013, Sell2014, Heckman2016}. 
In these galaxies, the outflow originate from a compact star-forming region with effective radius of several hundred parsecs. 
\citet{Heckman2016} found that the maximum outflow velocity ($v_{02}$, 10--2500 \kms) 
correlates with SFR (0.1--600 \sfrunit) and star formation rate density (SFRD, 0.1--5000 \sfrunit\,kpc$^{-2}$).  
J0916a also lies on the velocity--SFR relationship, with a fast outflow ($v_{02}\sim1600$ \kms\ for \oiii) and an intense star formation activity (SFR $= 990$ \sfrunit) similar with the most extreme starbursts in the sample of \citet{Sell2014} and \citet{Heckman2016}. 
Thus we can not rule out the possibility that star formation activity contributes to the fast outflow in J0916a, since the size of the star-forming region has not yet been constrained based on the current observations. 


\section{Conclusion}
\label{sec:conclusion}
In the spectroscopic follow-up observations of \akari-selected FIR-bright optically-faint objects, one ULIRG, J0916a, indicates signatures of an extremely strong outflow in its emission line profiles. Both of high- and low-ionization potential lines, e.g., \oiii\ and \oii, show large velocity dispersions and shifts in relative to the stellar absorption lines.
The velocity dispersions and shifts correspond to one of the fastest outflow among ULIRGs\,/\,DOGs at $0.3<z<1.0$, and are comparable to the obscured quasars at $z\sim2$. 
After the correction for a maximally possible contribution from an unresolved core component, the long-slit spectroscopy 2D image suggests that the outflow could extend to radius of 4 kpc. 
However, the co-existence of the strong outflow and vigorous starburst (SFR = 990 \sfrunit) suggests that the star formation has not yet been quenched by the outflow. 
The $L_{\rm bol}^{\rm \scriptscriptstyle AGN}$ of J0916a estimated from the SED fitting is only 5\,\%--10\,\% of $L_{\rm bol}^{\rm \scriptscriptstyle AGN}$ estimated from extinction-corrected \oiii\ luminosity, which leads to a large uncertainty in determinating the status of the AGN hidden in this ULIRG. 
Further observations are required in order to reveal the properties and origins of the outflow, e.g., 
hard X-ray observation to directly detect the intrinsic AGN radiation, 
spectroscopy observation with higher spatial and spectral resolution to determine the electron density in the outflowing gas, 
integral-field spectroscopy observation to investigate the structure of the outflow, 
and 
sub-millimeter observation to determine whether the warm ionized outflow really affect the cold molecular gas reservoirs. 

\begin{ack}
This research is based on data collected at Subaru Telescope, which is operated by the National Astronomical Observatory of Japan.
This research is based on observations with AKARI, a JAXA project with the participation of ESA. 
This publication makes use of data products from the Wide-field Infrared Survey Explorer, which is a joint project of the University of California, Los Angeles, and the Jet Propulsion Laboratory\,/\,California Institute of Technology, funded by the National Aeronautics and Space Administration.
Funding for SDSS-III has been provided by the Alfred P. Sloan Foundation, the Participating Institutions, the National Science Foundation, and the U.S. Department of Energy Office of Science. The SDSS-III web site is http://www.sdss3.org/.
SDSS-III is managed by the Astrophysical Research Consortium for the Participating Institutions of the SDSS-III Collaboration including the University of Arizona, the Brazilian Participation Group, Brookhaven National Laboratory, Carnegie Mellon University, University of Florida, the French Participation Group, the German Participation Group, Harvard University, the Instituto de Astrofisica de Canarias, the Michigan State\,/\,Notre Dame\,/\,JINA Participation Group, Johns Hopkins University, Lawrence Berkeley National Laboratory, Max Planck Institute for Astrophysics, Max Planck Institute for Extraterrestrial Physics, New Mexico State University, New York University, Ohio State University, Pennsylvania State University, University of Portsmouth, Princeton University, the Spanish Participation Group, University of Tokyo, University of Utah, Vanderbilt University, University of Virginia, University of Washington, and Yale University. 
MA is supported by JSPS KAKENHI (17H06129). 
YT is supported by Grant-in-Aid for JSPS Research Fellow (Grant No.18J01050). 
KI is supported by Program for Establishing a Consortium for the Development of Human Resources in Science and Technology, Japan Science and Technology Agency (JST) and JSPS KAKENHI (18K13584).
\end{ack}

\end{document}